\pdfoutput=1
\documentclass[11pt]{article}
\usepackage{amsmath,amsthm,amssymb,amscd}
\usepackage{hyperref}
\usepackage{geometry}
\usepackage{tikz-cd}
\usepackage{booktabs} % For better table aesthetics
\usepackage{graphicx} % For including images
\usepackage{multirow}
\usepackage{tikz}
\usetikzlibrary{calc,positioning,3d}
\usepackage[normalem]{ulem} % cross through
\usepackage{imakeidx}
\usepackage{makecell}
\usepackage{diagbox}
\usepackage{booktabs}
\usepackage{threeparttable}
\makeindex

\usepackage{caption}
\DeclareCaptionLabelFormat{adja-page}{\hrulefill\\}
\usepackage{subfig}

\theoremstyle{definition}

\title{Commutative Algebra Modeling of Mutation and Disease Association}

\author{
	JunJie Wee$^{1}$%
	\thanks{JunJie Wee and Mushal Zia share first authorship.},  
	Mushal Zia$^{1}$\footnotemark[1],
	Faisal Suwayyid$^{1,2}$, 
	Hongsong Feng$^{3}$, 
	Yuta Hozumi$^{1,4}$,
	and \\ Guo-Wei Wei$^{1,5,6}$%
	\thanks{Corresponding author: Guo-Wei Wei (guowei.wei@uga.edu).}\\[6pt]
	\parbox{0.85\textwidth}{
		\centering
		\footnotesize
		$^1$Department of Mathematics, Michigan State University, 
		East Lansing, MI 48824, USA.\\
		$^2$Department of Mathematics, King Fahd University of Petroleum and Minerals, 
		Dhahran 31261, KSA.\\
		$^3$Department of Mathematics and Statistics, University of North Carolina at Charlotte, 
		Charlotte, NC 28223, USA.\\
		$^4$School of Mathematics, Georgia Institute of Technology, 
		Atlanta, GA, USA.\\
		$^5$Department of Mathematics, University of Georgia, 
		Athens, GA 30602, USA.\\
		$^6$Department of Biochemistry and Molecular Biology, University of Georgia, 
		Athens, GA 30602, USA.
	}
}

\date{}

\begin{document}
	\maketitle
	 {\bf Abstract.}
		Genetic mutations can lead to disease by disrupting molecular interactions, altering protein stability, inducing misfolding, and promoting protein aggregation. Although existing predictive models have achieved promising performance, they often lack mechanistic interpretability and fail to fully incorporate the underlying physical and chemical principles governing mutation-induced effects. Moreover, current approaches typically address disease association, protein stability changes, and solubility alterations as separate prediction tasks rather than integrating them within a unified framework.  In this study, we introduce a multiscale commutative algebra modeling framework for predicting disease-mutation association. Leveraging persistent Stanley-Reisner theory, we extract algebraic invariants to build two predictive models: Commutative Algebra Neural Network (CANet) and Commutative Algebra Gradient Boosting Tree (CATree). The proposed framework is applied to three biologically connected tasks: disease-associated mutation identification, mutation-induced protein stability change prediction, and mutation-induced protein solubility change classification. Across six benchmark tasks, CANet and CATree consistently achieve leading predictive performance, with up to 7.5\% relative improvement over existing methods. These results demonstrate that commutative algebra provides a unified, interpretable, and generalizable modeling strategy for mutation-disease association prediction.

{\bf Relevance to Life Sciences.}
Proteins are the fundamental molecular machines of life, carrying out nearly all essential cellular processes, including catalysis, signaling, transport, structural support, and gene regulation. Their biological functions are  determined by their amino acid sequences. This work studies how different mutations can alter these sequences, thereby impairing normal cellular function and  causing human diseases, such as cancer, neuro-degenerative disorders, cardiovascular diseases, metabolic syndromes, and many inherited genetic conditions.

%\clearpage
%\section*{Significance}
%Commutative algebra offers foundational algebraic tools for algebraic geometry and algebraic number theory, yet it is rarely connected with data science and artificial intelligence. We introduce, for the first time, commutative algebra deep learning, adapting this abstract algebra for quantitative modeling and prediction. Our framework provides an interpretable neural architecture to capture mutation-induced structural and electrostatic changes. This novel approach reveals elusive connections between genetic diseases and critical mutations. Unlike existing models that lack mechanistic insight and generalizability, our approach unifies disease association, protein stability, and solubility prediction within a single model. Across six benchmark datasets, the proposed approach consistently outperforms other state-of-the-art methods, establishing an unprecedented paradigm for mathematically grounded, explainable AI.
	
 {\bf Mathematical Content.}
Commutative algebra offers foundational algebraic tools for algebraic geometry and algebraic number theory, yet it is rarely connected with data science and artificial intelligence. We introduce, for the first time, a commutative algebra deep learning framework, %adapting this abstract algebra 
for mutation-disease association  modeling and prediction. Persistent facet prime ideals and persistent $f$- and $h$-vectors are extracted from molecular structures to build the framework. We illustrate the interpretability of these algebraic invariants in capturing mutation-induced structural and electrostatic changes.

{\bf Keywords.}
Genetic disease, mutation, protein solubility, protein  stability, commutative algebra

{\bf MSC codes.}
92D99, 13P25

	\tableofcontents
\newpage
 
 %\linenumbers
 
\section{Introduction}

Genetic mutations, particularly missense variants, can disrupt protein structure, solubility, and stability, thereby contributing to disease through altered folding, impaired molecular interactions, and aggregation. These molecular perturbations are implicated in various diseases, including Alzheimer's disease \cite{hurley2023familial}, Parkinson's disease \cite{funayama2023molecular}, amyotrophic lateral sclerosis \cite{nakaya2018amyotrophic}, cancer \cite{chen2022mutant}, and metabolic syndromes \cite{zhang2013y328c,zhang2011silico}. Membrane proteins are especially important in this context because they account for approximately 20-30\% of human genes and serve as targets for more than half of small-molecule drugs, yet they constitute less than 0.5\% of experimentally resolved structures, largely because their essential roles in transport, signaling, adhesion, and cell entry are coupled with substantial structural instability outside lipid bilayers \cite{pires2020mcsm}. More broadly, over four million missense variants have been identified in the human genome, but only about 2\% have been clinically classified as pathogenic or benign \cite{karczewski2020mutational}. The remaining variants are often designated as variants of uncertain significance. Traditional approaches to variant discovery, including whole-genome sequencing, linkage analysis, and genome-wide association studies, are time-consuming and resource-intensive \cite{marian2020clinical}. They require extensive data generation, complex filtering, and manual curation. Functional validation using cell-based assays or animal models can take weeks to months, and many variants remain unresolved even after substantial experimental effort \cite{molotkov2024making}, creating a bottleneck for mechanistic interpretation and therapeutic development.

Existing predictive methods for identifying disease-associated mutations, such as PON-P2 \cite{niroula2015pon}, SIFT \cite{ng2003sift}, PolyPhen \cite{adzhubei2010polyphen}, and CADD \cite{kircher2014cadd}, provide efficient alternatives to labor-intensive site-directed mutagenesis experiments. However, many of these models have limited interpretability and may oversimplify the structural and physicochemical mechanisms through which mutations alter protein function. In addition, limited training data in previous studies, such as BORODA-TM \cite{popov2019prediction}, may lead to model overfitting, where strong validation performance does not necessarily translate into reliable generalization on blind test cases.

Protein stability is directly related to protein function because mutation-induced destabilization can perturb folding, conformational dynamics, molecular recognition, and downstream biological activity. Computational models for predicting mutation-induced protein stability changes have benefited from larger datasets, which enable the use of deep learning architectures, graph-based models, and multi-task learning algorithms. For example, mutDDG-SSM \cite{li2024prediction} integrates graph attention networks with gradient boosting trees to improve stability prediction. Similarly, TNet-MMP-2 \cite{cang2017topologynet} employs a multi-task, multi-channel framework to predict disease-associated mutations in globular proteins using protein stability data. Despite these advances, many models developed in the past decade, including STRUM \cite{quan2016strum}, use modern machine learning techniques and large datasets to uncover hidden relationships among protein sequence, structure, and stability, but still provide limited mechanistic interpretability.

Protein solubility is another essential determinant of protein function and disease. Changes in solubility can shift proteins away from their normally folded soluble states toward soluble oligomeric intermediates or insoluble aggregates. This relationship is particularly important in neurodegenerative diseases. In Alzheimer's disease, for example, amyloid beta peptides can aggregate through soluble oligomeric intermediates that are strongly associated with synaptic dysfunction, while larger insoluble deposits may participate in the broader aggregation process \cite{haass2007soluble}. This biological connection between soluble and insoluble protein states provides a clear rationale for studying mutation-induced solubility changes together with disease-associated mutations. Efforts to predict how mutations affect protein solubility have led to the development of several computational tools. A comprehensive review \cite{vihinen2020solubility} highlighted notable approaches, including CamSol \cite{sormanni2015camsol}, PON-Sol \cite{yang2016pon}, SODA \cite{paladin2017soda}, and Solubis \cite{van2016solubis}. Building on these efforts, PON-Sol2 \cite{yang2021pon} introduced an expanded dataset and used gradient boosting techniques to improve sequence-based solubility prediction. However, the overall predictive performance, quantified by the normalized Correct Prediction Ratio, remains suboptimal, indicating the need for more effective and interpretable modeling strategies.

Taken together, these observations reveal a critical gap: there is no unified framework that can jointly address disease-associated mutations, protein stability changes, and solubility alterations while preserving mechanistic interpretability. Essential physical interactions, including hydrogen bonding, van der Waals interactions, hydrophobic effects, salt bridges, and electrostatics, play central roles in maintaining wild-type protein function and in driving disease-causing molecular perturbations \cite{sun2022electrostatics,stefl2013molecular,zhang2013rational}. Therefore, there is an increasing need for an explainable artificial intelligence framework that connects predictive performance with interpretable molecular mechanisms underlying genetic diseases.

In this work, we introduce multiscale commutative algebra as a novel interpretable algebraic representation for capturing mutation-induced changes in intrinsic physical and chemical interactions. Commutative algebra is deeply rooted in algebraic geometry and number theory \cite{miller2005combinatorial,eisenbud2013commutative}, but its use in data science and machine learning remains limited. Here, we show that commutative algebra theory provides a powerful tool for representing intrinsic physical interactions and electrostatic shifts in three-dimensional protein structures. In particular, the proposed representation can characterize mutation-induced disruptions to hydrogen bonds, salt bridges, and other local interaction patterns, thereby providing an explainable framework for understanding the molecular basis of genetic diseases.

By integrating multiscale commutative algebra embeddings with auxiliary physical features and ESM-2 transformer features \cite{lin2023evolutionary}, we propose two learning models: Commutative Algebra Neural Network (CANet) and Commutative Algebra Gradient Boosting Tree (CATree). These models are designed to address disease-associated mutation identification, mutation-induced protein stability prediction, and mutation-induced solubility classification within a unified computational framework. The multiscale commutative algebra utilizes the persistent Stanley-Reisner theory  (PSRT) introduced by Suwayyid and Wei \cite{suwayyid2025persistent} to bridge commutative algebra and  data analysis. PSRT analyzes point cloud data through the evolution of square-free monomial ideals associated with simplicial complexes under filtration and produces computable algebraic invariants, including persistent facet prime ideals, $f$-vectors, $h$-vectors, and persistent graded Betti numbers. Across six benchmark tasks, the proposed commutative algebra-based models consistently outperform existing approaches, achieving up to 7.5\% relative improvement in predictive performance.

\section{Methods and Models}

In this section, we outline our multiscale commutative algebra embedding framework, data collection, and feature generation for wild-type and mutant protein structures. Computational settings for CANet and CATree are also provided along with the details on auxiliary features, ESM-2 descriptors and performance metrics.

\subsection{Persistent Stanley-Reisner Theory}
\label{subsec:PSRT}

Persistent Stanley-Reisner theory (PSRT) is a novel framework for analyzing the shape of data by leveraging tools from algebraic topology and commutative algebra. It encodes point cloud data as simplicial complexes, which are combinatorial structures built from vertices, edges, triangles, and higher-dimensional simplices. These complexes capture both topological and combinatorial features inherent in the data \cite{suwayyid2025persistent}. A filtration process is then applied to track the evolution and persistence of these features across multiple spatial or geometric scales. This approach introduces algebraic invariants such as persistent $h$-vectors, $f$-vectors, graded Betti numbers, and facet prime ideals, thereby providing a new algebraic perspective within the broader framework of topological data analysis.

\subsubsection{Persistent Stanley-Reisner Structures Over  Filtration}

Let \(\Delta\) be a simplicial complex on a finite vertex set
\(V=\{x_1,\dots,x_n\}\), and let
\(f:\Delta\to\mathbb{R}\) be a real-valued face function assigning a weight
to each face of \(\Delta\), such that \(f(\tau)\leq f(\sigma)\) whenever
\(\tau\subseteq\sigma\). The filtration of \(\Delta\) induced by \(f\) is
the family of subcomplexes
\begin{equation}\label{eq:filtered-complex}
    \widetilde{f}:=\left(\Delta^t\right)_{t\in\mathbb{R}},
    \quad\text{where}\quad
    \Delta^t:=
    \left\{
        \sigma\in\Delta
        \,\middle|\,
        f(\sigma)\leq t
    \right\}.
\end{equation}

This defines an increasing filtration in the sense that
\[
\Delta^s \subseteq \Delta^t \quad \text{whenever } s \le t.
\]

Fix a base field \(k\), and consider the standard polynomial ring
\[
S := k[x_1, \dots, x_n],
\]
graded by total degree. For each \(t \in \mathbb{R}\), associate to the filtered subcomplex \(\Delta^t\) the \emph{Stanley-Reisner ideal}
\begin{equation}\label{eq:sr-evolution}
	I^t := I(\Delta^t) := \left\langle x_{i_1} \cdots x_{i_r} \,\middle|\, \{x_{i_1}, \dots, x_{i_r}\} \notin \Delta^t \right\rangle \subseteq S,
\end{equation}
and its corresponding \emph{Stanley-Reisner ring}
\[
k[\Delta^t] := S / I^t.
\]
As the filtration parameter \(t\) increases, the subcomplex \(\Delta^t\) expands, resulting in a decreasing sequence of ideals:
\[
I^{s} \supseteq I^{t} \quad \text{for all } s \le t.
\]
Hence, the filtration of complexes induces a descending filtration of monomial ideals:
\[
I^{t_0} \supseteq I^{t_1} \supseteq I^{t_2} \supseteq \cdots \quad \text{for } t_0 < t_1 < t_2 < \cdots.
\]

Denote $\mathcal{F}(\Delta^t)$ as the set of all facets in the filtered subcomplex $\Delta^t$. Each ideal \(I^t\) admits a canonical decomposition as the intersection of prime monomial ideals associated to the facets of \(\Delta^t\):
\[
I^t = \bigcap_{\sigma \in \mathcal{F}(\Delta^t)} P_\sigma, \quad \text{where} \quad P_\sigma := (x_i \mid x_i \notin \sigma) \subseteq S.
\]
These primes \(P_\sigma\) are referred to as the \emph{facet prime ideals} at level \(t\). Let \(\mathcal{P}^t := \{ P_\sigma \mid \sigma \in \mathcal{F}(\Delta^t) \}\) denote the collection of all facet prime ideals at time \(t\).

To analyze their structure dimensionally, we stratify the collection \(\mathcal{P}^t\) by face dimension. For each \(i \ge 0\), define
\[
\mathcal{P}_i^t := \left\{ P_\sigma \in \mathcal{P}^t \,\middle|\, \dim(\sigma) = i \right\},
\]
so that
\[
\mathcal{P}^t = \bigsqcup_{i = 0}^{\dim(\Delta^t)} \mathcal{P}_i^t.
\]

In analogy with persistent homology, we introduce the notion of persistent facet prime ideals and their associated Betti invariants.

A facet prime ideal \(P_\sigma \in \mathcal{P}_i^t\) is said to \emph{persist} to level \(t' > t\) if \(P_\sigma \in \mathcal{P}_i^{t'}\). The set of such persistent \(i\)-dimensional facet prime ideals is
\[
\mathcal{P}_i^{t,t'} := \mathcal{P}_i^t \cap \mathcal{P}_i^{t'}.
\]

The \emph{facet persistence Betti number} \(\beta_i^{t,t'}\) is defined as
\[
\beta_i^{t,t'} := \left| \mathcal{P}_i^t \cap \mathcal{P}_i^{t'} \right|,
\]
which counts the number of \(i\)-dimensional facet prime ideals common to both filtration levels \(t\) and \(t'\).

The collection \(\{\beta_i^{t,t'}\}_{i,t,t'}\) provides a graded combinatorial signature of the evolving prime decomposition of the Stanley-Reisner ideals across the filtration. These values quantify the persistence of prime components and serve as an algebraic counterpart to topological barcodes in persistent homology.

\subsubsection{Persistent Graded Betti Numbers of Stanley-Reisner Rings}

Let \(k\) be a field and \(S = k[x_1, \dots, x_n]\) the standard graded polynomial ring. For each filtration level \(t \in \mathbb{R}\), the Stanley-Reisner ring \(k[\Delta^t] := S / I^t\) inherits a natural \(\mathbb{Z}\)-graded structure as an \(S\)-module. As such, it admits a minimal graded free resolution of the form
\begin{equation}\label{eq:min-free-res}
	\cdots 
	\longrightarrow
	\bigoplus_{j} S(-j)^{\beta_{i,j}(k[\Delta^t])}
	\longrightarrow
	\cdots 
	\longrightarrow
	\bigoplus_{j} S(-j)^{\beta_{0,j}(k[\Delta^t])}
	\longrightarrow
	k[\Delta^t]
	\longrightarrow
	0,
\end{equation}
where \(S(-j)\) denotes the graded free module \(S\) shifted in degree \(j\), and the exponents
\begin{equation}\label{eq:Betti-numbers-def}
	\beta_{i,j}(k[\Delta^t]) := \dim_k \operatorname{Tor}^S_i(k[\Delta^t], k)_j
\end{equation}
are the \emph{graded Betti numbers} of \(k[\Delta^t]\), measuring the number of minimal generators in homological degree \(i\) and internal degree \(j\).

To relate these algebraic invariants to the topology of the simplicial complex, we recall that for any subset \(W \subseteq V\), the \emph{induced subcomplex} on \(W\) is defined by
\begin{equation}\label{eq:induced-subcomplex}
	\Delta_W^t := \left\{ \sigma \in \Delta^t \,\middle|\, \sigma \subseteq W \right\}.
\end{equation}

Hochster’s formula expresses the graded Betti numbers of \(k[\Delta^t]\) in terms of the reduced simplicial homology of these subcomplexes:
\begin{equation}\label{eq:Hochster-general}
	\beta_{i,j+i}(k[\Delta^t])
	=
	\sum_{\substack{W \subseteq V \\ |W| = j+i}} 
	\dim_k \widetilde{H}_{j-1}(\Delta_W^t; k),
\end{equation}
where \(\widetilde{H}_{j-1}(\Delta_W^t; k)\) denotes the \((j-1)\)-st reduced simplicial homology group over \(k\).

In particular, Hochster’s formula can be reformulated in terms of the (non-reduced) Betti numbers of induced subcomplexes. For each integer \(i \ge 0\), the following identities hold:
\begin{align}
	\beta_{i,i+1}(k[\Delta^t]) &= \sum_{\substack{W \subseteq V \\ |W| = i+1}} \left( \beta_0(\Delta_W^t) - 1 \right), \label{eq:Hochster-j1} \\
	\beta_{i,i+j}(k[\Delta^t]) &= \sum_{\substack{W \subseteq V \\ |W| = i+j}} \beta_{j-1}(\Delta_W^t), \quad \text{for all } j \ge 2, \label{eq:Hochster-j2}
\end{align}
where \(\Delta_W^t\) denotes the subcomplex of \(\Delta^t\) induced on the vertex subset \(W \subseteq V\), and \(\beta_r(\Delta_W^t)\) denotes the \(r\)-th Betti number of \(\Delta_W^t\) with coefficients in \(k\). 

Equation~\eqref{eq:Hochster-j1} quantifies the combinatorial deviation from connectivity in induced subcomplexes of cardinality \(i+1\), while Equation~\eqref{eq:Hochster-j2} captures contributions from higher-dimensional homology classes across subsets of size \(i+j\).

\vspace{1em}
\noindent
We now define a persistent refinement of these Betti numbers that tracks how graded components persist across filtration levels.
Let \(t \le t'\). The \emph{persistent graded Betti number} \(\beta_{i,i+j}^{t,t'}(k[\Delta])\) is defined by
\begin{equation}\label{eq:persistent-Betti}
\beta_{i,i+j}^{t,t'}(k[\Delta])
:=
\sum_{\substack{W \subseteq V \\ |W| = i+j}}
\dim_k \operatorname{im}\left(
\iota_{j-1}^{t,t'}(W):
\widetilde{H}_{j-1}(\Delta_W^t;k)
\longrightarrow
\widetilde{H}_{j-1}(\Delta_W^{t'};k)
\right),
\end{equation}
where \(\iota_{j-1}^{t,t'}\) denotes the homomorphism on reduced homology induced by the inclusion \(\Delta_W^t \hookrightarrow \Delta_W^{t'}\).

This definition yields a multigraded generalization of classical persistent Betti numbers. It captures not only the persistence of topological features in the filtration but also refines this by the combinatorial structure of their supports.

When \(i+j=|V|\), the only subset \(W\subseteq V\) satisfying
\(|W|=i+j\) is \(W=V\). Hence, the persistent graded Betti number reduces to
\[
\beta_{i,|V|}^{t,t'}(k[\Delta])
=
\dim_k \operatorname{im}\left(
\widetilde{H}_{|V|-i-1}(\Delta^t;k)
\longrightarrow
\widetilde{H}_{|V|-i-1}(\Delta^{t'};k)
\right),
\]
recovering the persistent Betti number of homological degree
\(|V|-i-1\). More generally, the family
\(\{\beta_{i,i+j}^{t,t'}(k[\Delta])\}_{i,j}\) encodes a richer multiscale
invariant that interpolates between algebraic and topological persistence.

\subsubsection{Persistent \texorpdfstring{\(f\)}{f}- and \texorpdfstring{\(h\)}{h}-Vectors}

Let \((\Delta^t)_{t \in \mathbb{R}}\) be a filtration of a finite \((d-1)\)-dimensional simplicial complex \(\Delta\), induced by some face function \(f: \Delta \to \mathbb{R}\). For each fixed level \(t\), the complex \(\Delta^t\) consists of those faces \(\sigma \in \Delta\) with \(f(\sigma) \le t\). At each level \(t\), one may associate the classical combinatorial invariants of face counts and their derived quantities.

The \emph{\(f\)-vector} of \(\Delta^t\) is defined as
\[
f(\Delta^t) = \left(f_{-1}^t, f_0^t, f_1^t, \dots, f_{d-1}^t\right),
\]
where \(f_i^t\) denotes the number of \(i\)-dimensional faces in \(\Delta^t\), and \(f_{-1}^t = 1\) by convention. From this, one obtains the associated \emph{\(h\)-vector} \(h(\Delta^t) = (h_0^t, \dots, h_d^t)\), given by the linear transformation:
\begin{equation}\label{eq:h-vector}
	h_m^t = \sum_{j=0}^{m} 
	\binom{d - j}{m - j} 
	(-1)^{m-j} f_{j-1}^t,
	\quad \text{for } m = 0, \dots, d.
\end{equation}
This transformation is invertible, with inverse relation
\begin{equation}\label{eq:f-vector}
	f_{m-1}^t = \sum_{i=0}^m 
	\binom{d - i}{m - i} h_i^t,
	\quad \text{for } m = 0, \dots, d.
\end{equation}

To extend these invariants to the persistent setting, one replaces the classical Betti numbers with the persistent graded Betti numbers \(\beta_{i,j}^{t,t'}\) defined over filtration levels \(t \le t'\) \cite{suwayyid2025persistent}. This allows for a multiscale combinatorial interpretation of how face structures persist across filtration levels.

Let \((\Delta^t)_{t \in \mathbb{R}}\) be a filtration of a simplicial complex \(\Delta\), and let \(d = \dim(\Delta) + 1\). The \emph{persistent \(h\)-vector} between levels \(t \le t'\) is defined as
\begin{equation}\label{eq:persistence_h}
	h_m^{t,t'} := \sum_{j=0}^{m}
	\binom{n - d + m - j - 1}{m - j}
	\left(
	\sum_{i=0}^{j} (-1)^i \beta_{i,j}^{t,t'}
	\right),
	\quad \text{for } m = 0, \dots, d,
\end{equation}
where \(\beta_{i,j}^{t,t'}\) denotes the persistent graded Betti numbers of \(k[\Delta]\) over \([t,t']\).

The corresponding \emph{persistent \(f\)-vector} is then defined via the inverse transformation:
\begin{equation}\label{eq:persistence_f}
	f_{m-1}^{t,t'} := \sum_{i=0}^m
	\binom{d - i}{m - i} \, h_i^{t,t'},
	\quad \text{for } m = 0, \dots, d.
\end{equation}

These persistent vectors capture how the combinatorial structure of \(\Delta\) evolves through the filtration, blending face enumeration with homological persistence. In contrast to the classical static \(f\)- and \(h\)-vectors, their persistent counterparts reflect the dynamic appearance and disappearance of faces and their relations across multiple scales, providing a richer algebraic-combinatorial invariant for topological data analysis.

\subsection{Commutative Algebra Modeling}
\subsubsection{Multiscale Commutative Algebra Embedding}
\label{sec:Vectorization}

\paragraph{Element-specific and site-specific atom sets.}

In this work, multiscale commutative algebra embeddings are constructed from element-specific and site-specific atom sets. This representation simplifies protein geometry while preserving key mutation-induced physical interactions through algebraic invariants.

\begin{figure}[!htb]
	\centering
	\includegraphics[width=.75\textwidth]{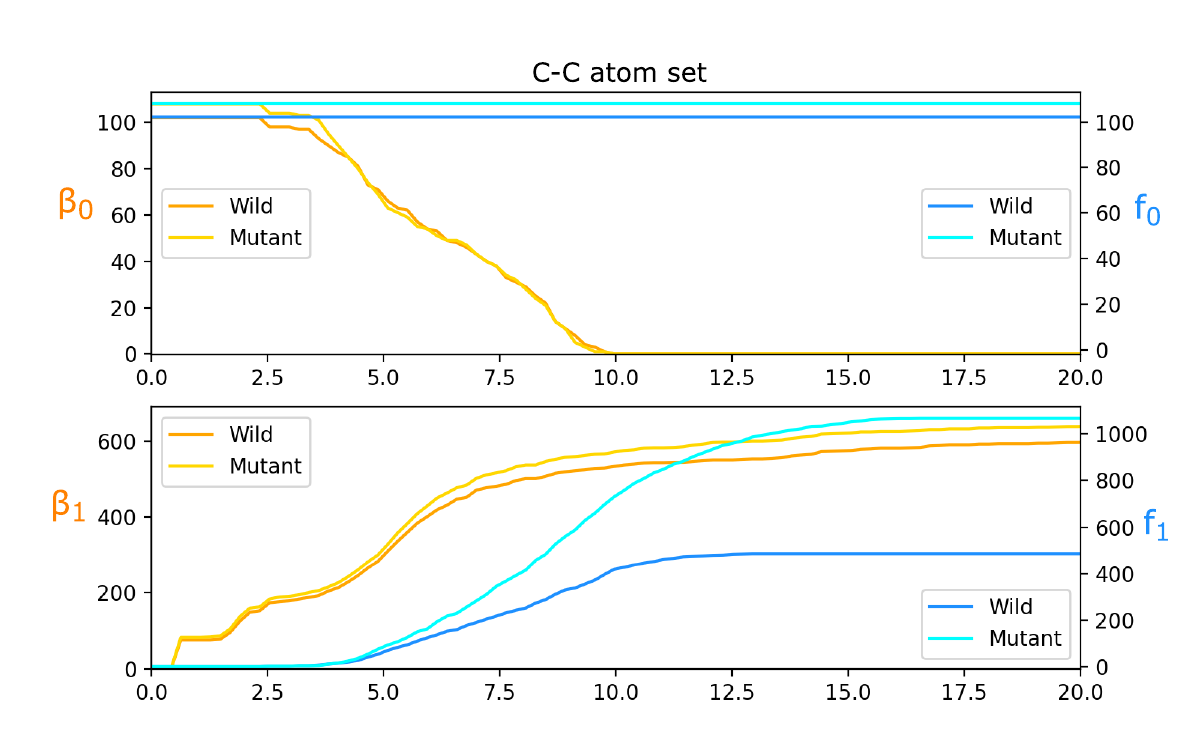}
	\caption{Top panel: An illustration of dimension-0 commutative algebraic descriptors for C-C interactions from VR complexes using ${\rm D}_{\mathrm{mod}}$-based filtration at residue 204 mutation site from Q to W in protein ID: 213133708 of the PON-Sol2 dataset. Bottom panel: An illustration of dimension-1 commutative algebraic descriptors for C-C interactions from VR complexes using ${\rm D}_{\mathrm{mod}}$-based filtration at residue 204 mutation site from Q to W in protein ID: 213133708. Left and right axes: Dim-0 and dim-1 facet prime ideals, $\beta_0$ and $\beta_1$, and $f$-vectors, $f_0$ and $f_1$, respectively.}
	\label{fig:CC}
\end{figure}

\begin{figure}[!htb]
	\centering
	\includegraphics[width=.75\textwidth]{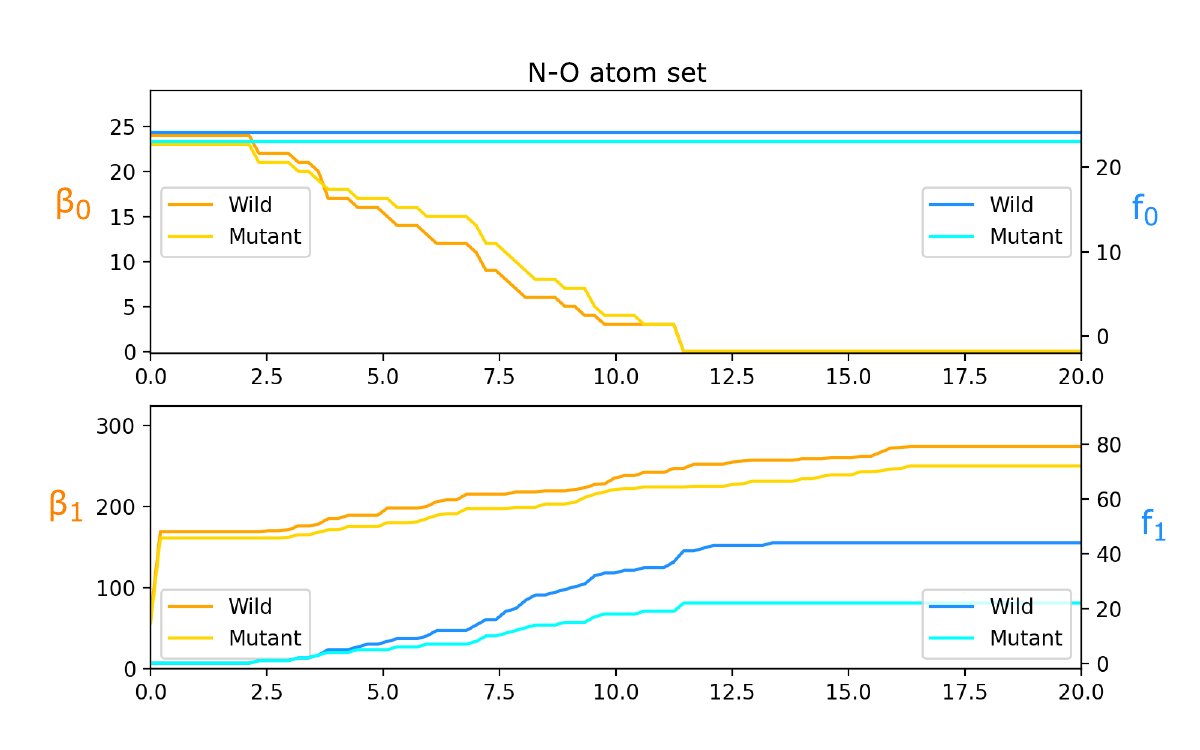}
	\caption{Top panel: An illustration of dimension-0 commutative algebraic descriptors for N-O interactions from VR complexes using ${\rm D}_{\mathrm{mod}}$-based filtration at residue 204 mutation site from Q to W in protein ID: 213133708. Bottom panel: An illustration of dimension-1 commutative algebraic descriptors for N-O interactions from VR complexes using ${\rm D}_{\mathrm{mod}}$-based filtration at residue 204 mutation site from Q to W in protein ID: 213133708.}
	\label{fig:NO}
\end{figure}

Atoms in a 3D protein structure are first partitioned into site-specific sets. The mutation-site atom set is denoted by $\mathcal{A}_m$, while the mutation-neighborhood atom set within a cutoff radius $r$ is denoted by $\mathcal{A}_{mn}(r)$. Atoms are further classified into element-specific subsets $\mathcal{A}_{\varepsilon}$, where $\varepsilon \in \{C, N, O\}$. This construction yields nine pairwise element combinations between mutation-site atoms and mutation-neighborhood atoms, and these combinations encode distinct interaction types. For example, $\mathcal{A}_C \cap \mathcal{A}_m$ paired with $\mathcal{A}_C \cap \mathcal{A}_{mn}(r)$ captures hydrophobic C-C interactions, whereas $\mathcal{A}_N \cap \mathcal{A}_m$ paired with $\mathcal{A}_O \cap \mathcal{A}_{mn}(r)$ captures hydrophilic N-O interactions, including hydrogen-bond and salt-bridge related contacts. Figure~\ref{fig:CC} illustrates the hydrophobic C-C interactions, while Figure~\ref{fig:NO} illustrates the hydrophilic N-O interactions associated with hydrogen-bond and salt-bridge related contacts.

%%The element-specific and site-specific atom sets characterize different interaction types in the commutative algebra embeddings. 

To focus on interactions between mutation-site atoms and their surrounding atomic environment, we modify the standard Euclidean distance so that the interactions within the mutation site or within the mutation neighborhood are excluded. For atoms $a_i$ and $a_j$, the modified Euclidean distance ${\rm D}_{\mathrm{mod}}$ is defined as,
\begin{equation}
	{\rm D}_{\mathrm{mod}}(a_i, a_j) =
	\begin{cases}
		\infty, & \text{if } a_i,a_j \in \mathcal{A}_{m} \text{ or } a_i,a_j \in \mathcal{A}_{mn}(r),\\
		{\rm DE}(a_i, a_j), & \text{otherwise}.
	\end{cases}
\end{equation}
Here, ${\rm DE}(\cdot, \cdot)$ denotes the Euclidean distance between two atoms. Therefore, finite filtration values arise only from cross interactions between mutation-site atoms and mutation-neighborhood atoms.

\subsubsection{Commutative Algebra Feature Generation}

Using the element-specific and site-specific atom subsets, commutative algebraic features are generated from both wild-type and mutant protein structures using Vietoris-Rips (VR) and Alpha complexes. The VR complex forms simplices from atoms whose pairwise distances fall within a selected filtration threshold, while the Alpha complex is derived from Delaunay triangulation and retains simplices enclosed within a specified radius. Moreover, the modified Euclidean distance ${\rm D}_{\mathrm{mod}}$ is used to construct the VR complexes, whereas the standard Euclidean distance is used for the Alpha complexes.

In this study, atoms within $16\,\text{\AA}$ of the mutation site are selected to represent the local mutation environment. The filtration parameter ranges from $1\,\text{\AA}$ to $12\,\text{\AA}$ in increments of $0.4\,\text{\AA}$. At each filtration value, the dimension-0 and dimension-1 facet prime ideals and $f$-vectors are computed. These features are concatenated to construct the multiscale commutative algebra embedding. The wild-type features, mutant features, and their corresponding differences are included for both the facet-ideal and $f$-vector representations.

As illustrative examples, Figure~\ref{fig:CC} and Figure~\ref{fig:NO} show the facet prime ideals and $f$-vector curves generated from protein ID: 213133708 in the PON-Sol2 dataset. Figure~\ref{fig:CC} uses the atom sets $\mathcal{A}_C \cap \mathcal{A}_m$ and $\mathcal{A}_C \cap \mathcal{A}_{mn}(r)$ to construct VR complexes with ${\rm D}_{\mathrm{mod}}$-based filtration, revealing hydrophobic C-C interactions. Similarly, Fig.~\ref{fig:NO} uses the atom sets $\mathcal{A}_N \cap \mathcal{A}_m$ and $\mathcal{A}_O \cap \mathcal{A}_{mn}(r)$ to construct VR complexes with ${\rm D}_{\mathrm{mod}}$-based filtration, revealing hydrophilic N-O interactions associated with hydrogen-bond and salt-bridge related contacts.

\subsubsection{CANet Architecture and CATree Model Hyperparameters}

The deep neural network in CANet consists of six hidden layers, each containing 15,000 neurons, and produces either a regression output or a classification label depending on the prediction task. For protein stability change prediction, the model is trained for 200 epochs using a batch size of 32 and a learning rate of 0.001. For protein solubility change classification, the model is also trained for 200 epochs with the same learning rate, while the batch size is set to 50.

In this study, we also use a Commutative Algebra gradient boosting Tree (CATree) model to handle small datasets such as M546, which contains a limited number of mutation samples for training deep neural networks. The gradient boosting tree (GBT) models are implemented using the Python \texttt{scikit-learn} package (v1.3.2) \cite{scikit-learn}. GBT models are well suited for small to medium-sized datasets because of their robustness, relative insensitivity to hyperparameter settings, and ease of implementation. The algorithm builds an ensemble of weak learners, where each individual tree contributes to the final prediction. Although a single weak learner may have limited predictive power, the ensemble reduces overall error by combining the predictions of many trees.

The resulting commutative algebraic descriptors and ESM-2 transformer features are used as input features for CATree. The CATree hyperparameters used in this study are listed in \autoref{table:CATree-parameters}.

\begin{table}[htb!] \small \centering \begin{tabular}{c c c c } \hline No. of estimators & Max depth & Min. sample split & Learning rate\\ 20000 & 7 & 3 & 0.05\\ \hline Max features & Subsample size & Repetition &\\ Square root & 0.4& 10 times & \\ \hline \end{tabular} \caption{CATree model hyperparameters.} \label{table:CATree-parameters} \end{table}

\subsection{Data and Vectorization}

\subsubsection{Data Collection} 
To rigorously assess the performance of CANet and CATree, we curated a diverse set of benchmark datasets spanning three core tasks related to mutation-induced effects on protein structure and function: Disease-associated mutations, protein stability changes, and protein solubility changes. 

\begin{figure}[!htb]
	\centering
	\includegraphics[width=1\textwidth]{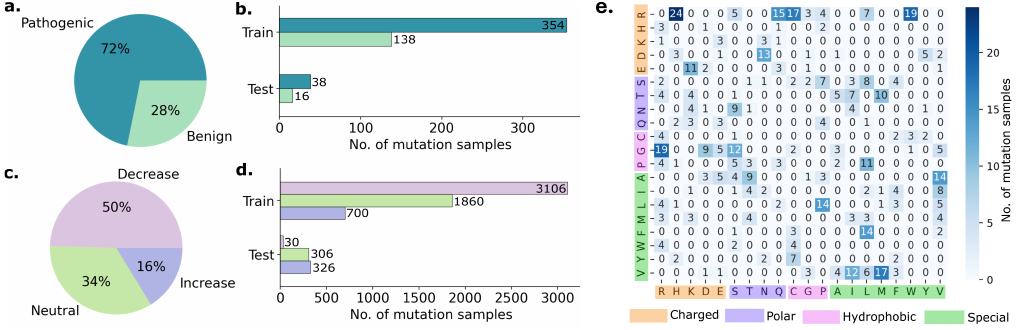}
	\caption{Distribution and classification of mutation samples across benchmark datasets. \textbf{a.} Composition of the M546 dataset, showing the number of pathogenic and benign mutations. \textbf{b.} Allocation of M546 mutations into training and blind test sets. \textbf{c.} Categorization of PON-Sol2 mutations by solubility effect: decreased, increased, or unchanged. \textbf{d.} Distribution of solubility classes across training and test sets in the PON-Sol2 blind test. \textbf{e.} Heatmap showing the number of mutation samples for different amino acid substitutions in the M546 dataset. The y-axis indicates the original amino acid residue, and the x-axis indicates the substituted (mutant) residue. Residues are grouped by biochemical properties: charged, polar, hydrophobic, and special. Color intensity reflects the frequency of each mutation pair.}
	\label{fig:dist}
\end{figure}

\paragraph{M546 dataset} The M546 dataset originated from the BORODA-TM \cite{popov2019prediction} study, which focuses on disease-associated mutations in transmembrane proteins with known 3D structures. It comprises of 392 pathogenic and 154 benign point mutations across 64 transmembrane proteins. The breakdown of pathogenic and benign point mutations for M546 are summarized in Figure~\ref{fig:dist}a and Figure~\ref{fig:dist}b. Notably, 492 mutations are used for the 10-fold cross-validation while the remaining 54 samples are used in a blind test prediction. These mutations were curated from UniProtKB/Swiss-Prot and mapped to structural data from the Human Transmembrane Proteome (HTP) database. The dataset emphasizes mutations located within transmembrane helices, which are particularly relevant for understanding membrane protein dysfunction in disease.

It is important to note that a recent study from mCSM \cite{pires2020mcsm} revealed a marked discrepancy between cross-validation (MCC = 0.87, F1-Score = 0.96) and blind test performance from BORODA-TM (MCC = 0.46, F1-Score = 0.78), suggesting potential overfitting and underscoring the need for robust generalization. In addition, mCSM \cite{pires2020mcsm} successfully processed only 539 samples because seven mutations involving two residues in PDB entry 4ZWJ \cite{kang2015crystal} could not be mapped owing to missing residues at the mutation sites. 

To further characterize the M546 dataset, we analyzed the distribution of amino acid substitutions across four biochemical mutation categories. The 20 amino acids were grouped as charged residues (R, H, K, D, E), polar residues (S, T, N, Q), hydrophobic residues (C, G, P), and special residues (A, I, L, M, F, W, Y, V), following the grouping used in our mutation-type analysis. Figure~\ref{fig:dist}e shows the number of mutation samples for each original and substituted amino acid pair. The distribution is uneven across residue pairs, indicating that the M546 dataset has limited diversity for several mutation types. For example, among the 35 charged-to-special mutations, 19 correspond to R-to-W substitutions, suggesting that some mutation categories are dominated by a small number of residue-level substitutions.

\paragraph{S2648 and S350 dataset} The S2648 dataset \cite{dehouck2009fast} contains 2,648 single-point mutations across 132 proteins, annotated with experimentally measured changes in protein stability. It serves as a comprehensive benchmark for regression-based prediction of mutation-induced stability changes. The S350 dataset is a curated subset of S2648, consisting of 350 mutations in 67 proteins, and is commonly used for blind testing and comparative evaluation of predictive models.

\paragraph{PON-Sol2 dataset} The PON-Sol2 dataset \cite{yang2021pon} includes 6,328 mutations across 77 proteins, annotated with protein solubility changes: increased, decreased, or neutral (unchanged). The distribution of PON-Sol2 mutation samples by solubility effect and their allocation to the training and blind-test sets are summarized in Figure~\ref{fig:dist}c and Figure~\ref{fig:dist}d, respectively. It was developed to support machine learning-based prediction of mutation-induced solubility changes. The dataset addresses class imbalance and includes both experimentally validated and literature-curated variants. It is widely used for multi-class classification tasks in protein solubility prediction. 

%%%\newpage
%%%\section{Datasets}

Overall, the M546 and PON-Sol2 datasets provide stringent benchmarks for evaluating the robustness of CANet and CATree under conditions of pronounced class imbalance. M546 comprises a binary classification task, whereas PON-Sol2 involves multi-class prediction. Both datasets include variants curated from literature and validated experimentally and the extent of class imbalance across these datasets is detailed in Figure~\ref{fig:dist}.

\subsubsection{Performance Metrics}

For all classification tasks, positive predictive value (PPV), negative predictive value (NPV), specificity, and sensitivity are calculated from the true-positive (TP), true-negative (TN), false-positive (FP), and false-negative (FN) counts for each solubility class:
\begin{equation}
{\rm PPV}=\frac{TP}{TP+FP},
\qquad
{\rm NPV}=\frac{TN}{TN+FN}.
\end{equation}
\begin{equation}
{\rm Specificity}=\frac{TN}{TN+FP},
\qquad
{\rm Sensitivity}=\frac{TP}{TP+FN}.
\end{equation}

For mutation-induced protein stability change prediction, Pearson correlation coefficient (PCC) and root mean squared error (RMSE) are used to evaluate CANet and CATree:
\begin{equation}
{\rm PCC}
=
\frac{\sum_{i=1}^{n}(x_i-\bar{x})(y_i-\bar{y})}
{\sqrt{\sum_{i=1}^{n}(x_i-\bar{x})^2
		\sum_{i=1}^{n}(y_i-\bar{y})^2}},
\qquad
{\rm RMSE}
=
\sqrt{\frac{1}{n}\sum_{i=1}^{n}(y_i-\hat{y}_i)^2}.
\end{equation}
Here, $x_i$ and $y_i$ are the values of variables $x$ and $y$ for the $i$th sample, while $\bar{x}$ and $\bar{y}$ are their corresponding means. In the RMSE calculation, $y_i$ and $\hat{y}_i$ denote the true and predicted values, respectively, and $n$ is the number of samples.

For mutation-induced protein solubility change prediction, the accuracy score, also referred to as the correct prediction ratio (CPR), and the generalized squared correlation ${\rm GC}^2$ are used to evaluate CANet and CATree:
\begin{equation}
{\rm Accuracy}
=
\frac{1}{N}\sum_i z_{ii},
\qquad
{\rm GC}^2
=
\frac{1}{N(K-1)}
\sum_{i,j}
\frac{(z_{ij}-e_{ij})^2}{e_{ij}}.
\end{equation}
Here, $K$ is the number of classes, $N$ is the number of samples, and $z_{ij}$ is the number of samples from class $i$ predicted as class $j$. Defining $x_i=\sum_j z_{ij}$ as the number of inputs from class $i$ and $y_j=\sum_i z_{ij}$ as the number predicted as class $j$, the expected value of the $(i,j)$th confusion-matrix entry is
\begin{equation}
e_{ij}=\frac{x_i y_j}{N}.
\end{equation}
Because the mutational samples are imbalanced across the three solubility classes, the values are normalized to provide more reliable performance measurements.

\subsubsection{Auxiliary Descriptors}

To complement the commutative algebraic descriptors in CANet, auxiliary descriptors capturing essential chemical and physical properties are divided into atom-level and residue-level features.

\paragraph{Atom-level Features}

Atom-level features include atomic types, partial charges, van der Waals radii, Coulomb energies, solvation energies, and solvent-excluded surface areas. Atoms are grouped by element (C, N, O, S, H, heavy atoms, and all atoms), spatial proximity (mutation site, within 10\AA, and all atoms), and mutation state (wild type, mutant, and their difference).

Partial-charge descriptors include the sum and absolute sum of charges for each atomic group, producing 126 features. Coulomb energies are calculated as
\begin{equation}
C_i=\sum_{j\ne i}\frac{q_iq_j}{r_{ij}},
\end{equation}
where $q_i$ and $q_j$ are partial charges and $r_{ij}$ is the interatomic distance. Five element types and both signed and absolute energies are considered, producing 90 features.

Van der Waals energies are calculated using the Lennard-Jones potential:
\begin{equation}
V_i
=
\sum_{j\ne i}
\epsilon
\left[
\left(\frac{r_i+r_j}{r_{ij}}\right)^{12}
-
2\left(\frac{r_i+r_j}{r_{ij}}\right)^6
\right],
\end{equation}
where $\epsilon=1$ and a cutoff distance of 40\AA\ is applied, yielding 45 features. Solvent-excluded surface areas are calculated using ESES \cite{liu2017eses}, and the atomic areas within each group are summed to produce 63 features.

\paragraph{Residue-Level Features}

Residue-level descriptors include the amino acid composition within 10\AA\ of the mutation site. Amino acids are divided into hydrophobic, polar, positively charged, negatively charged, and special-case groups. The counts and percentages of residues in each group produce 10 features.

The sum, mean, and variance of residue volume, surface area, molecular weight, and hydropathy score provide an additional 12 features. The p$K_a$ values of seven ionizable residues, ASP, GLU, ARG, LYS, HIS, CYS, and TYR, are calculated using PROPKA \cite{bas2008very}. Differences between the wild-type and mutant p$K_a$ values, referred to as p$K_a$ shifts, are used to characterize mutation-induced changes in ionization states.

\subsubsection{ESM-2 Transformer Features}
To incorporate evolutionary context into the algebraic framework, sequence embeddings generated by the ESM-2 transformer model are integrated \cite{lin2023evolutionary}. Following the success of protein transformer models such as ESM \cite{rives2021biological} and ProtTrans \cite{vaswani2017attention, Devlin2019BERTPO}, ESM-2 was trained on 250 million protein sequences using a masked language modeling objective. The model consists of 34 transformer layers and 650 million parameters. For each input sequence of length $L$, ESM-2 produces a contextualized representation of size $1280 \times L$ at each layer, excluding special tokens. The representation from the final, 34th layer is extracted and averaged across the sequence length dimension, yielding a fixed 1280-dimensional vector.

These embeddings are combined with the multiscale commutative algebra embeddings as input features for CANet and CATree. This integration combines transformer-derived evolutionary sequence information with interpretable structure-based algebraic descriptors, improving generalization across diverse mutation prediction tasks.

\section{Results} 

\subsection{Overview of CANet Workflow} 

\begin{figure}[!htb]
	\centering
	\includegraphics[width=.9\textwidth]{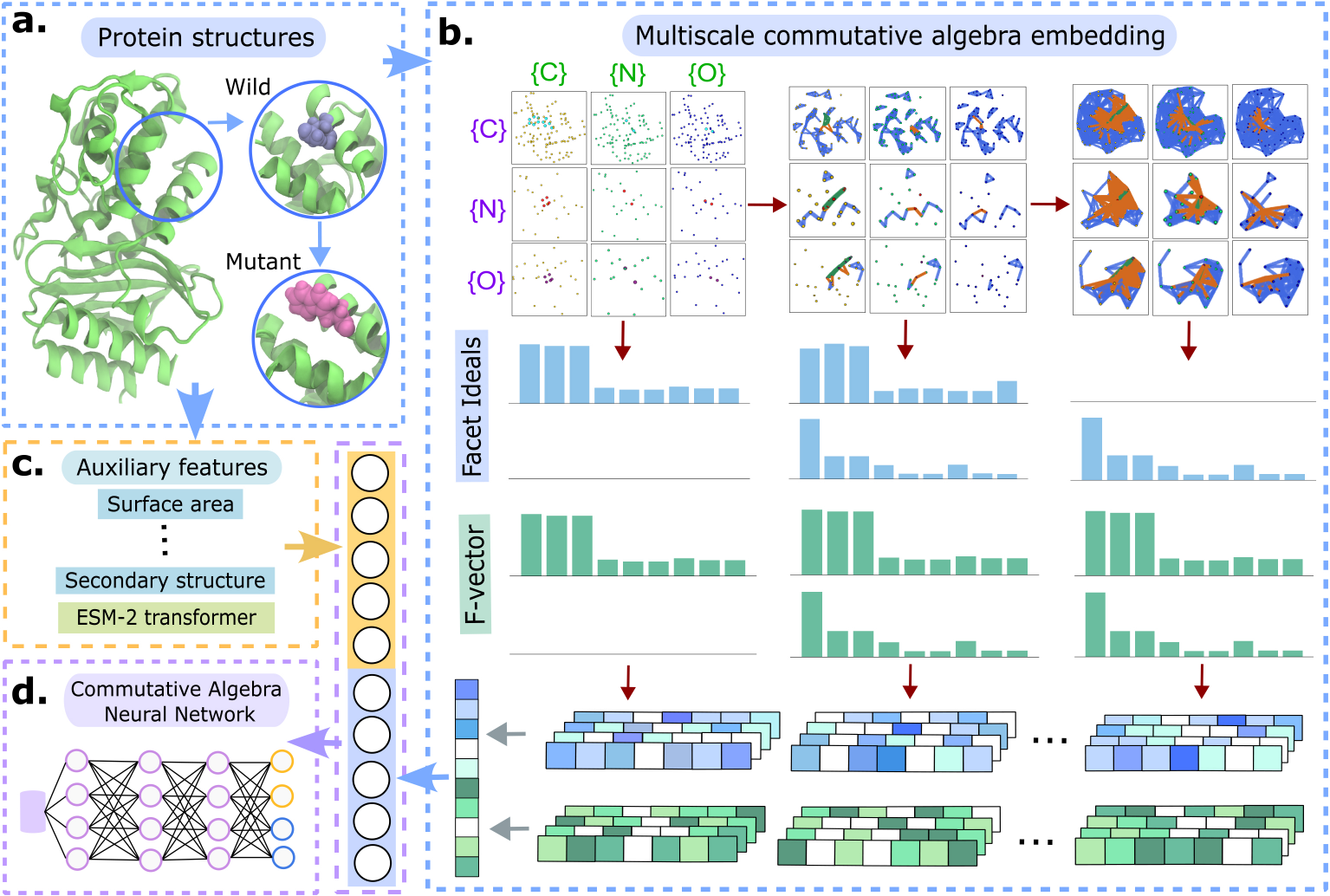}
	\caption{Illustration of commutative algebra neural network (CANet) workflow. \textbf{a.} 3D protein structures obtained from the PDB. Mutant proteins are generated from the Jackal software \cite{xiang2002jackal}. \textbf{b.} Mutational-site and its local neighborhood atom subsets are extracted from both the wild-type and mutant structures to form element specific subcomplexes. Multiscale commutative algebra embedding is performed to generate the persistent facet prime ideals and persistent $f$-vector curves. \textbf{c.} Auxiliary features such as surface area, secondary structure and ESM-2 transformer-based features are also generated. \textbf{d.} Commutative algebra features are concatenated with auxiliary and ESM-2 transformer-based features to form a long feature vector. Features are then fed into the downstream CANet model. The hyperparameters of CANet are optimized. Colors of dotted frames and arrows indicate workflows in different modules: \textbf{a.} and \textbf{b.} Commutative algebra-based module (blue), \textbf{c.} Auxiliary and ESM-2 transformer-based module (orange), \textbf{d.} CANet module (purple).}
	\label{fig:concept}
\end{figure}

Figure~\ref{fig:concept} summarizes the proposed CANet workflow for three related prediction tasks: identifying disease-associated mutations, predicting mutation-induced protein stability changes, and predicting protein solubility changes. Because these tasks involve different biological endpoints but share common mutation-driven structural and physicochemical mechanisms, a unified and interpretable representation is needed.

The workflow begins with wild-type protein structures collected from available datasets, and corresponding mutant structures are generated using Jackal software \cite{xiang2002jackal}. This procedure models the local structural effect of the target mutation while preserving the overall wild-type structural context. From each wild-type and mutant structure, local atom subsets around the mutation site are extracted and separated into element-specific subcomplexes. Persistent Stanley-Reisner theory (PSRT) \cite{suwayyid2025persistent} is then used to construct multiscale commutative algebra embeddings from these subcomplexes. In particular, persistent facet prime ideals and persistent $f$-vector curves are computed under structural filtrations to encode mutation-associated changes in local geometry and interaction patterns.

\begin{figure}[!htbp]
	\centering
	\includegraphics[width=.8\textwidth]{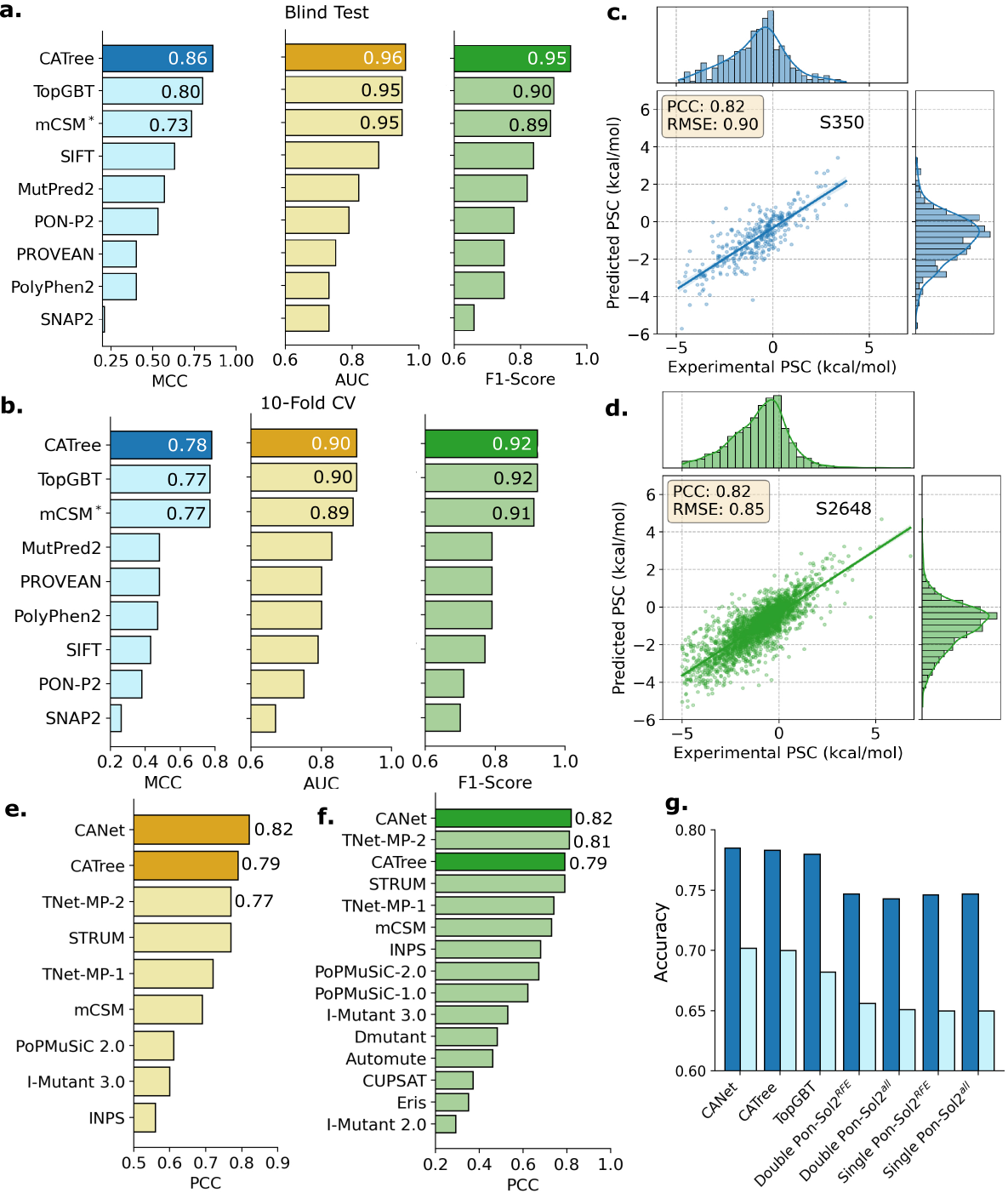}
	\caption{Illustration of commutative algebra model performance in predicting disease-related mutations, protein stability changes and protein solubility changes upon mutation. \textbf{a.} Blind test performance of CATree with existing state-of-the-art models \cite{pires2020mcsm} in predicting disease-associated mutations. \textbf{b.} 10-fold cross-validation performance of CATree in predicting disease-associated mutations. \textbf{c.} Comparison of experimental PSC with predicted ones from CANet for S350 dataset.  \textbf{d.} Comparison of experimental protein stability changes (PSC) with predicted ones from CANet for S2648 dataset.  \textbf{e.} Performance of CANet and CATree for S2648 dataset compared to existing state-of-the-art models \cite{cang2017topologynet,worth2011sdm,quan2016strum}. \textbf{f.} Performance of CANet and CATree for S350 dataset compared to existing state-of-the-art models \cite{cang2017topologynet,worth2011sdm,quan2016strum}. \textbf{g.} Accuracy scores of CANet and CATree for mutation-induced protein solubility change classification compared with existing state-of-the-art models \cite{wee2024integration,yang2021pon}. Dark blue bars represent the raw accuracy scores, whereas light blue bars represent the corresponding normalized accuracy scores.}
	\label{fig:results}
\end{figure}

In parallel, auxiliary structural features, including solvent-accessible surface area and secondary structure annotations, are generated together with ESM-2 transformer-based sequence embeddings. The commutative algebra features, auxiliary features, and ESM-2 features are concatenated into a unified representation and passed to the downstream learning model for training and prediction. In addition to CANet, a commutative algebra gradient boosting tree (CATree) model is applied for tasks with smaller datasets, where tree-based learning can offer stronger robustness under limited sample size.

The interpretable design of this framework allows commutative algebraic features to be traced back to mechanistic changes in local protein environments, including electrostatic shifts, altered interaction patterns, and hydrogen bond disruptions. As a biologically grounded explainable artificial intelligence framework, CANet connects predictive performance with mutation-induced structural mechanisms and supports robust prediction across diverse protein systems.

\subsection{Identifying Disease-Associated Mutations}

In this study, we first assess the predictive capability of commutative algebra models in identifying disease-associated mutations. We conduct random 10-fold cross-validation and an independent blind test using the M546 dataset, which contains 392 pathogenic and 154 benign mutations from 63 transmembrane proteins \cite{popov2019prediction}. For example, the glycine-to-arginine mutation at position 380 of the fibroblast growth factor receptor 3 (FGFR3) gene is a primary cause of achondroplasia and is therefore labeled as pathogenic. A total of 492 mutations were used for training in the cross-validation procedure, while the remaining 54 mutations were reserved for blind testing to evaluate model generalization.

Because of the limited dataset size and the small number of available transmembrane protein mutation samples for deep learning, existing models \cite{popov2019prediction,pires2020mcsm} have shown signs of overfitting, with strong cross-validation performance but weaker generalization in blind testing. To benchmark performance, we also applied TopGBT \cite{wee2024integration}, a persistent homology-based model, to the M546 dataset. As shown in Figure~\ref{fig:results}a, CATree outperformed TopGBT in blind test prediction, achieving a Matthews correlation coefficient (MCC) of 0.86, corresponding to a 7.5\% relative improvement over TopGBT. CATree also achieved an area under the ROC curve (AUC) of 0.96 and an F1-score of 0.95, demonstrating strong robustness and accuracy in predicting disease-associated mutations.

Consistent with the blind test results, CATree also performed strongly in 10-fold cross-validation, as shown in Figure~\ref{fig:results}b. The model achieved an MCC of 0.78, corresponding to a 1.3\% relative improvement over TopGBT. CATree also achieved an AUC of 0.90 and an F1-score of 0.92, showing predictive performance comparable to TopGBT and higher than the other evaluated state-of-the-art models.

\subsection{Mutation-Induced Protein Stability Change Prediction}

Perturbations in protein stability are a critical molecular mechanism underlying functional disruption in many genetic diseases. These changes can alter folding dynamics, interaction interfaces, and overall structural integrity, thereby contributing to pathogenic phenotypes. To evaluate the predictive performance of our models in this context, we used the S2648 dataset, which contains 2,648 mutation samples across 131 protein structures annotated with stability changes ($\Delta\Delta G$, kcal/mol) \cite{dehouck2009fast}. Model performance was assessed through two complementary tasks. First, we performed 5-fold cross-validation on the full S2648 dataset to evaluate general predictive performance. Second, we conducted targeted testing on the S350 dataset, a curated benchmark subset of S2648 designed for stability prediction evaluation. Predictive accuracy was quantified using the Pearson correlation coefficient (PCC) and root mean squared error (RMSE), which measure linear agreement and absolute deviation between predicted and experimental values, respectively.

\begin{figure}[!htb]
	\centering
	\includegraphics[width=1\textwidth]{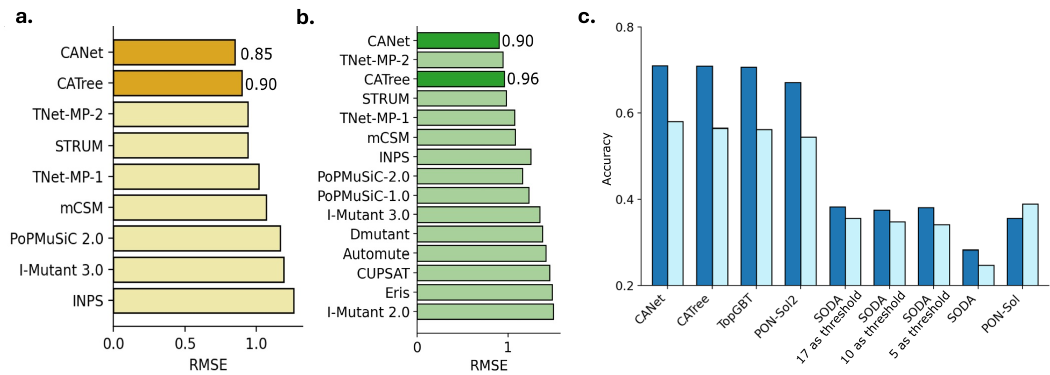}
	\caption{\textbf{a.} Root mean squared error (RMSE) values for the S2648 5-fold cross-validation task. \textbf{b.} RMSE values for the S350 test set prediction task. Lower RMSE indicates better predictive performance. \textbf{c.} The predictive accuracy of CANet and CATree compared with various existing state-of-the-art models in the blind test set prediction for the mutation-induced protein solubility change dataset.}
	\label{fig:newfig}
\end{figure}

Here, we evaluate CANet and the Commutative Algebra Gradient Boosting Tree (CATree) for predicting mutation-induced protein stability and solubility changes. For the S2648 protein stability dataset, CANet showed strong agreement between the predicted and experimentally measured protein stability changes, as illustrated in Figure~\ref{fig:results}d. In 5-fold cross-validation, CANet achieved a PCC of 0.82, outperforming all existing methods and corresponding to a 6.49\% relative improvement over TNet-MP-2, a leading topological convolutional neural network model \cite{cang2017topologynet}, as shown in Figure~\ref{fig:results}e. CANet also achieved the lowest RMSE of 0.85, representing a 9.6\% relative reduction compared with TNet-MP-2, as shown in Figure~\ref{fig:newfig}a. These results demonstrate the effectiveness of commutative algebra-based embeddings in capturing mutation-associated structural perturbations. 

On the S350 test set, CANet maintained strong predictive performance, with a PCC of 0.82 and an RMSE of 0.90, as illustrated in Figure~\ref{fig:results}c. Its PCC exceeded that of TNet-MP-2 by 1.23\%, while its RMSE remained lower than those of the existing state-of-the-art models, as shown in Figure~\ref{fig:results}f and Figure~\ref{fig:newfig}b, respectively. CATree also demonstrated competitive performance on both S2648 and S350, further supporting the effectiveness of the proposed commutative algebra features across different machine learning architectures.

For mutation-induced protein solubility change classification, CANet and CATree achieved normalized accuracies of 0.580 and 0.565, respectively, in the blind test. These results correspond to relative improvements of up to 6.4\% over the PON-Sol2 models and up to 3.2\% over TopGBT, as shown in Figure~\ref{fig:newfig}c.

Similarly, CATree also showed competitive performance, outperforming all existing models on the S350 test set except TNet-MP-2. In contrast to several existing methods, both CANet and CATree can process all mutation samples in the benchmark datasets, further supporting their reliability. For example, I-Mutant 3.0 evaluated only 2,636 of 2,648 samples in S2648 and 338 of 350 samples in S350, as reported in Table~\ref{table:PCC}. Moreover, Table~\ref{table:PCC} summarizes the performance of CANet, CATree, and existing methods, together with the number of mutation samples processed by each model. These results demonstrate that commutative algebra embeddings, particularly when integrated into CANet, provide robust and scalable representations for accurately predicting protein stability changes.

\subsection{Mutation-Induced Protein Solubility Change Classification}

Genetic mutations can alter protein solubility and thereby contribute to disease through aggregation, misfolding, and impaired cellular function. These effects are especially relevant to neurodegenerative disorders, metabolic syndromes, and cancer, where changes in soluble and insoluble protein states can influence disease progression. To evaluate mutation-associated solubility changes, we applied our commutative algebra-based models to the PON-Sol2 dataset \cite{yang2021pon}, which contains 6,328 mutation samples from 77 distinct proteins. Each sample is labeled as decreased, increased, or unchanged solubility. Among these mutations, 3,136 are associated with decreased solubility, 1,026 with increased solubility, and 2,166 with no observable change. The dataset is class imbalanced, with an approximate ratio of 1:0.34:0.69 and a predominance of decreased-solubility mutations, as shown in Figure~\ref{fig:dist}c.

\begin{figure}[!htb]
	\centering
	\includegraphics[width=0.7\textwidth]{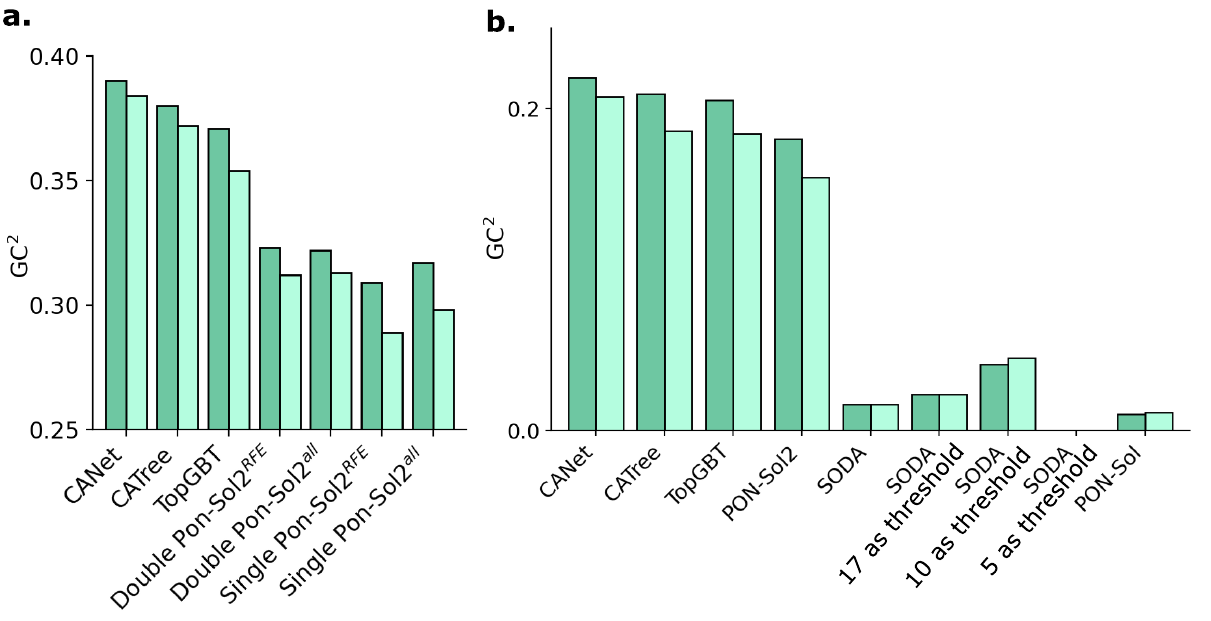}
	\caption{ \textbf{a.} The generalized squared correlation GC$^2$ scores in the 10-fold cross-validation. \textbf{b.} The GC$^2$ scores in the blind test prediction task.}
	\label{fig:GC2}
\end{figure}

To assess model performance, we conducted 10-fold cross-validation and an independent blind test. Figure~\ref{fig:dist}d shows the distribution of mutation samples across the training and test sets used in the blind test. In the cross-validation setting, CATree and CANet were compared with TopGBT \cite{wee2024integration}, a persistent homology-based model, and existing PON-Sol2 models \cite{yang2021pon}, which use feature selection strategies such as recursive feature elimination. Because this is a multi-class classification task, normalized accuracy was used to evaluate predictive performance, as shown in Figure~\ref{fig:results}g. CANet and CATree achieved normalized accuracies of 0.702 and 0.700, respectively, corresponding to relative improvements of up to 7.01\% over existing PON-Sol2 models and up to 2.93\% over TopGBT.

To further evaluate robustness, we used the generalized squared correlation (GC$^2$), which captures nonlinear dependencies between predicted and true labels. Benchmark comparisons based on (GC$^2$) are presented in Figure~\ref{fig:GC2}a and Figure~\ref{fig:GC2}b. Both CANet and CATree achieved higher (GC$^2$) scores than the competing models, indicating that commutative algebra-based representations improve generalization in class-imbalanced solubility classification tasks.

\begin{table}[!htb]
	\centering
	\small
	\setlength{\tabcolsep}{5pt}
	\caption{Comparison of CANet and CATree with existing methods for mutation-induced protein stability change prediction. PCC and RMSE are reported for the S2648 and S350 benchmark datasets.}
	\label{table:PCC}
	\begin{threeparttable}
		\begin{tabular*}{\textwidth}{@{\extracolsep{\fill}}lccc lccc@{}}
			\toprule
			\multicolumn{4}{c}{\textbf{S2648}} & \multicolumn{4}{c}{\textbf{S350}} \\
			\cmidrule(lr){1-4} \cmidrule(lr){5-8}
			\textbf{Method} & \textbf{$n^{d}$} & \textbf{PCC} & \textbf{RMSE} 
			& \textbf{Method} & \textbf{$n^{d}$} & \textbf{PCC} & \textbf{RMSE} \\
			\midrule
			CANet           & 2648 & \textbf{0.82} & \textbf{0.85} & CANet            & 350 & \textbf{0.82} & \textbf{0.90} \\
			CATree          & 2648 & 0.79          & 0.90          & TNet-MP-2        & 350 & 0.81          & 0.94          \\
			TNet-MP-2       & 2648 & 0.77          & 0.94          & CATree           & 350 & 0.79          & 0.96          \\
			STRUM$^{b}$     & 2647 & 0.77          & 0.94          & STRUM$^{b}$      & 350 & 0.79          & 0.98          \\
			TNet-MP-1       & 2648 & 0.72          & 1.02          & TNet-MP-1        & 350 & 0.74          & 1.07          \\
			mCSM$^{b,c}$    & 2643 & 0.69          & 1.07          & mCSM$^{b,c}$     & 350 & 0.73          & 1.08          \\
			PoPMuSiC2.0$^{b}$ & 2647 & 0.61        & 1.17          & INPS$^{b,c}$     & 350 & 0.68          & 1.25          \\
			I-Mutant3.0$^{b}$ & 2636 & 0.60        & 1.19          & PoPMuSiC2.0$^{b}$ & 350 & 0.67         & 1.16          \\
			INPS$^{b,c}$    & 2648 & 0.56          & 1.26          & PoPMuSiC 1.0$^{a}$ & 350 & 0.62        & 1.23          \\
			\multicolumn{4}{c}{}                                      & I-Mutant3.0$^{b}$ & 338 & 0.53          & 1.35          \\
			\multicolumn{4}{c}{}                                      & Dmutant$^{a}$     & 350 & 0.48          & 1.38          \\
			\multicolumn{4}{c}{}                                      & Automute$^{a}$    & 315 & 0.46          & 1.42          \\
			\multicolumn{4}{c}{}                                      & CUPSAT$^{a}$      & 346 & 0.37          & 1.46          \\
			\multicolumn{4}{c}{}                                      & Eris$^{a}$        & 334 & 0.35          & 1.49          \\
			\multicolumn{4}{c}{}                                      & I-Mutant 2.0$^{a}$ & 346 & 0.29         & 1.50          \\
			\bottomrule
		\end{tabular*}
		\begin{tablenotes}
			\footnotesize
			\item Existing results are referenced from TNet-MP-2~\cite{cang2017topologynet} unless otherwise stated.
			\item[$^{a}$] Results obtained from SDM~\cite{worth2011sdm}.
			\item[$^{b}$] Results obtained from STRUM~\cite{quan2016strum}.
			\item[$^{c}$] According to STRUM~\cite{quan2016strum}, the online-server PCC and RMSE values for the S350 task are 0.59 and 1.28 for INPS, and 0.70 and 1.13 for mCSM, respectively.
			\item[$^{d}$] Number of mutation samples successfully processed.
		\end{tablenotes}
	\end{threeparttable}
\end{table}

\section{Discussion}

After demonstrating the strong performance of CANet and CATree across disease-associated mutation identification, protein stability change prediction, and protein solubility change classification, we next examine how mutation-related factors influence predictive accuracy. In particular, we analyze the effects of mutation region, such as interior or surface locations, and mutation type, such as charged, polar, hydrophobic, or special residue substitutions, on model performance. These analyses help identify the biological contexts in which commutative algebra-based models perform strongly and where limitations may remain.

\subsection{Predictive Performance}

The stratified mutation-type analysis should be interpreted in light of the uneven amino acid substitution distribution in the M546 dataset. As shown in Figure~\ref{fig:newfig}d, several mutation categories are represented by relatively few residue-level substitutions, while some categories are dominated by specific substitution patterns, such as R-to-W mutations. This limited diversity may affect performance estimates for certain mutation groups and highlights the need for larger and more balanced transmembrane mutation datasets in future studies.

Beyond predictive accuracy, the algebraic structure of the proposed models provides a natural basis for interpretability. Commutative algebra features capture mutation-related changes in local geometry, electrostatic organization, and physical interaction patterns. These features can therefore connect model predictions with mechanistic changes in protein environments, including electrostatic shifts, hydrogen bond disruptions, and salt bridge formation. This interpretability supports the use of CANet and CATree as biologically grounded explainable artificial intelligence models for understanding disease-associated mutations and mutation-related changes in protein stability and solubility.

\begin{figure}[!htb]
	\centering
	\includegraphics[width=.8\textwidth]{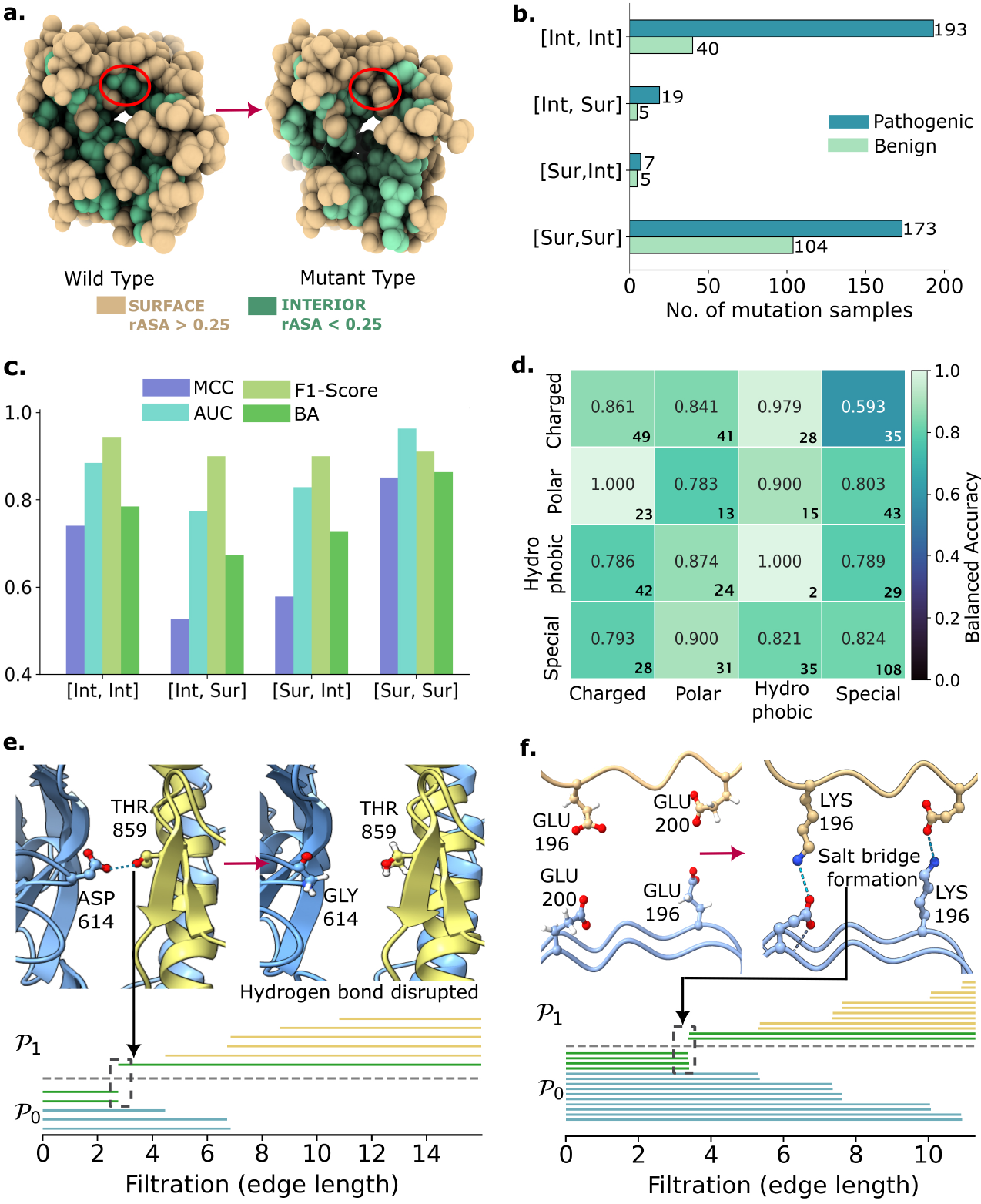}
	\caption{Electrostatic interaction analysis and mutation effects on protein structure and pathogenicity \cite{sun2022electrostatics}. \textbf{a.} Structural shift caused by the A38E mutation in human aquaporin 5, a membrane protein structure from PDB ID 3D9S, from the surface region to the interior region [Sur, Int]. \textbf{b.} Distribution of pathogenic and benign mutations in the M546 dataset across mutation region pairs. \textbf{c.} Balanced accuracy of CATree on the M546 dataset stratified by mutation region pair. \textbf{d.} CATree prediction performance grouped by amino acid type, with bold numbers indicating the sample count in each cell. \textbf{e.} Persistent facet prime ideals reveal a hydrogen bond interaction before its disruption by the D614G mutation in the SARS-CoV-2 spike protein structure, PDB ID 6VSB \cite{wrapp2020cryo}. The hydrogen bond is represented by the emergence of a green dimension-1 facet prime ideal after two green dimension-0 facet prime ideals stop persisting at 2.74~\AA.}
	\label{fig:discussion}
\end{figure}

\begin{figure}[!htb]
	\ContinuedFloat
	\centering
	\caption{Electrostatic interaction analysis and mutation effects on protein structure and pathogenicity, continued. \textbf{f.} Persistent facet prime ideals illustrate salt bridge formation in an amyloid fibril structure, PDB ID 7DWV~\cite{wang2021genetic}, following the E196K mutation associated with genetic prion disease. Dimension-0 facet prime ideals persist up to 3.35~\AA\ and 3.40~\AA\ in green, representing two N-O atom pairs. The emergence of green dimension-1 facet prime ideals at these distances marks salt bridge formation and reflects new electrostatic interactions introduced by lysine.}
\end{figure}

Overall, the results suggest that commutative algebra-based representations provide a unified and interpretable framework for mutation modeling. By combining multiscale algebraic descriptors with auxiliary structural features and ESM-2 transformer features, CANet and CATree achieve strong predictive performance across distinct but biologically connected tasks. At the same time, the stratified analyses show that model performance depends on dataset composition, mutation region, and mutation type. These observations emphasize both the promise of commutative algebra modeling and the importance of continued dataset expansion for improving generalization across diverse protein systems.

\paragraph{Impact of Mutation Region on Predictive Performance.}
We further examined prediction outcomes by categorizing mutation residues according to their structural regions. Relative accessible surface area (rASA), following the criteria in \cite{levy2010simple}, was used to classify wild-type and mutant residues as either surface-exposed (Sur) or interior (Int) \cite{bogan1998anatomy}. This classification produced four mutation-region categories: [Int, Int], [Int, Sur], [Sur, Int], and [Sur, Sur], representing the region assignments before and after mutation. Figure~\ref{fig:discussion}a illustrates a representative structural shift from the surface to the interior region, while Figure~\ref{fig:discussion}b shows the sample distribution across the four categories in the M546 dataset. Similar rASA-based mutation-region classifications have been used in previous studies to characterize protein mutation environments and interfaces~\cite{wee2024integration,chen2023topological,wang2020topology}.

CATree performance across these four categories is shown in Figure~\ref{fig:discussion}c. The [Int, Int] and [Sur, Sur] categories achieved balanced accuracies of 0.78 and 0.86, respectively, outperforming the mixed-region categories [Int, Sur] and [Sur, Int]. Similar trends were observed for MCC, AUC, and F1-score, indicating that mutations remaining within the same structural region are predicted more reliably than mutations associated with region changes. This difference may partly reflect the limited number of samples in the mixed-region categories, as shown in Figure~\ref{fig:discussion}b, which can reduce model generalization. These results suggest that future expansion of mixed-region mutation samples may improve the evaluation and predictive robustness of commutative algebra-based models for disease-associated mutations.

\paragraph{Impact of Mutation Type on Predictive Performance.}
We also assessed how biochemical mutation types influence CATree performance. The 20 canonical amino acids were grouped into charged, polar, hydrophobic, and special categories, as described in the dataset characterization section. Figure~\ref{fig:discussion}d shows the balanced accuracy of CATree across pairwise mutation-type combinations, with bold numbers indicating sample counts. CATree achieved strong performance for several mutation types, including charged-to-hydrophobic substitutions with a balanced accuracy of 0.979 from 28 samples, polar-to-charged substitutions with a balanced accuracy of 1.000 from 23 samples, and hydrophobic-to-polar substitutions with a balanced accuracy of 0.874 from 31 samples. These results suggest that commutative algebra features are sensitive to mutations that induce substantial electrostatic changes, alter side-chain polarity, or disrupt hydrophobic packing, which are common mechanisms in pathogenic variants.

Lower balanced accuracies were observed for charged-to-special substitutions, with a score of 0.593 from 35 samples, and polar-to-polar substitutions, with a score of 0.783 from 13 samples. These cases may involve more context-dependent physicochemical changes that are harder to separate using limited training data. As shown in Figure~\ref{fig:discussion}, 19 of the 35 charged-to-special mutations correspond to R-to-W substitutions. In transmembrane proteins, R-to-W mutations can be rare and structurally complex because they involve loss of positive charge and introduction of a bulky hydrophobic side chain, which may perturb membrane insertion, helix packing, or electrostatic anchoring \cite{molnar2016characterization}. Polar-to-polar mutations may also preserve hydrogen-bonding potential and side-chain polarity, producing weaker changes in local geometry or electrostatics. These observations highlight the importance of residue-level diversity and balanced mutation sampling when interpreting model performance across biochemical mutation categories.

\subsection{Commutative Algebra Enabled Explainable AI} 

Understanding the origin of genetic diseases requires a rational learning framework capable of interpreting the biological and structural mechanisms underlying pathogenic mutations. In this study, we proposed CANet as an eXplainable AI (XAI) model that leverages commutative algebra to represent protein structures and infer mutation-induced perturbations. CANet achieves interpretability through algebraic embeddings based on persistent Stanley-Reisner rings, which encode multiscale facet prime ideals and $f$-vector features derived from protein structural filtrations. These algebraic embeddings bridge computational biophysics and XAI by providing mechanistic insights into how specific mutations can lead to diseases and alter protein stability, solubility, and electrostatic interactions.

\begin{figure}[!htb]
	\centering
	\includegraphics[width=1\textwidth]{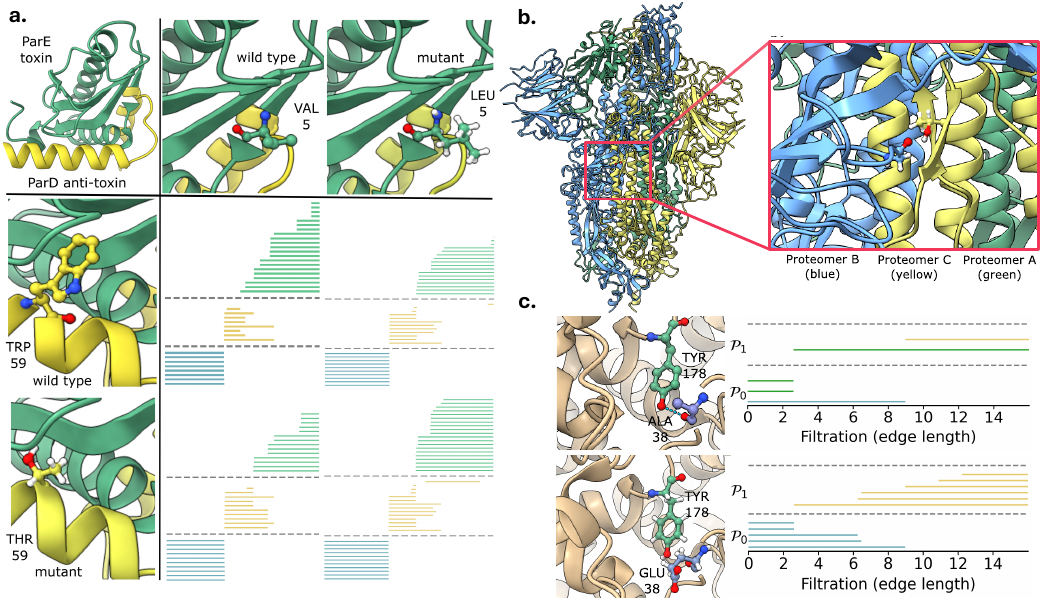}
	\caption{ \textbf{a.} Illustration of co-evolutionary mutations from the ParE toxin (green) and ParD anti-toxin (yellow) that expand future mutational pathways that sustain existing interactions while also forming new ones \cite{ding2022co}. The ParD anti-toxin W59 mutates into threonine, thereby potentially enabling V5 of ParE toxin to mutate into leucine. For each of the four scenarios among the wild and mutant toxin and anti-toxins, commutative algebra provides a set of persistent facet prime ideals to explain the intrinsic structural patterns and changes. \textbf{b.} Illustration of SARS-CoV-2 spike protein (PDBID: 6VSB \cite{wrapp2020cryo}) after D614G mutation and the region surrounding G614 (Proteomer B) and T859 (Proteomer C). Note that this region is outside the receptor binding domain (RBD) of SARS-CoV-2, but D614G is a popular mutation that enhanced the infectivity of SARS-CoV-2 during the COVID-19 pandemic \cite{plante2021spike}. \textbf{c.} Illustration of residue A38 before (top left) and after mutation (bottom left) and its corresponding persistent facet prime ideals for human aquaporins AQP5 (top and bottom right). Left panel: The residue A38 mutates from alanine (ALA) into a charged residue, glutamic acid (GLU). Right panel: The mutation changes the persistent facet prime ideals, which not only reflects the disruption of hydrogen bonds but also the O-O interactions between residue 178 and residue 38.}
	\label{fig:newfig2}
\end{figure}

The interpretability of CANet is demonstrated through several representative mutation examples, where commutative algebra reveals physical interactions such as hydrogen-bond disruption, salt-bridge formation, and electrostatic shifts induced by protein mutations. Fig.~\ref{fig:newfig2}a extends the interpretability of commutative algebra by examining co-evolutionary mutational effects, which produce distinct persistent facet barcodes under different scenarios. Figure~\ref{fig:newfig2}b shows the mutation region of D614G in the SARS-CoV-2 spike protein (PDBID: 6VSB) \cite{wrapp2020cryo}, while Figure~\ref{fig:newfig2}c illustrates the disruption of hydrogen bonds in human aquaporin AQP5 after the mutation of A38 to the charged residue E.

\paragraph{Protein Properties Captured by Commutative Algebra.}

To demonstrate the interpretability of CANet, we traced its commutative algebraic representations, specifically persistent facet prime ideals, back to biologically meaningful protein functionalities that govern electrostatic interactions and conformational stability. This analysis reveals that algebraic invariants serve as an eXplainable AI framework capable of elucidating the critical role of electrostatics in disease-associated mutations. In the SARS-CoV-2 spike protein, CANet identifies hydrogen bonds with average distances of approximately $3.0\AA$ between residues T859 and D614, as shown in Figure~\ref{fig:discussion}e. The D614G mutation, a defining feature of early viral evolution, disrupts this interaction and contributes to enhanced infectivity during the initial wave of the COVID-19 pandemic \cite{plante2021spike}. In addition, Figure~\ref{fig:newfig2}b highlights the mutation region associated with this structural change. In Figure~\ref{fig:discussion}e, two dimension-0 facet prime ideals (green bars) persist up to $2.74\AA$, corresponding to oxygen atoms from ASP614 and THR859. At this threshold, a dimension-1 ideal emerges, indicating hydrogen bond formation. This interaction is lost upon mutation, thereby altering viral fitness. A similar pattern is observed in the transmembrane protein human aquaporin AQP5. As shown in Figure~\ref{fig:newfig2}c, the charged mutation A38E disrupts hydrogen bonding between A38 and Y178. This is consistent with prior studies showing that A38E perturbs the local electrostatic environment, affects loop conformation, and ultimately alters the protein's tertiary and quaternary structure \cite{calvanese2018structural}. 

Furthermore, Figure~\ref{fig:discussion}f illustrates a salt bridge with an approximate distance of 3.40$\AA$ formed in the amyloid fibril structure following the E196K mutation, which is associated with genetic prion disease. The substitution introduces a positively charged lysine that engages in electrostatic and hydrogen bond interactions with glutamic acid, thereby stabilizing the fibril through salt-bridge formation. As an extension, algebraic invariants can also be applied to extract interpretable co-evolutionary structural changes that strongly influence protein mutational pathways \cite{ding2022co}. In this regard, Figure~\ref{fig:newfig2}a demonstrates distinct persistent facet barcodes for different co-evolutionary mutation scenarios, further supporting the interpretability of commutative algebra in capturing mutation-driven structural variation.

\paragraph{Structural Motifs Captured by Commutative Algebra.}

\begin{figure}[!htb]
	\centering
	\includegraphics[width=.95\textwidth]{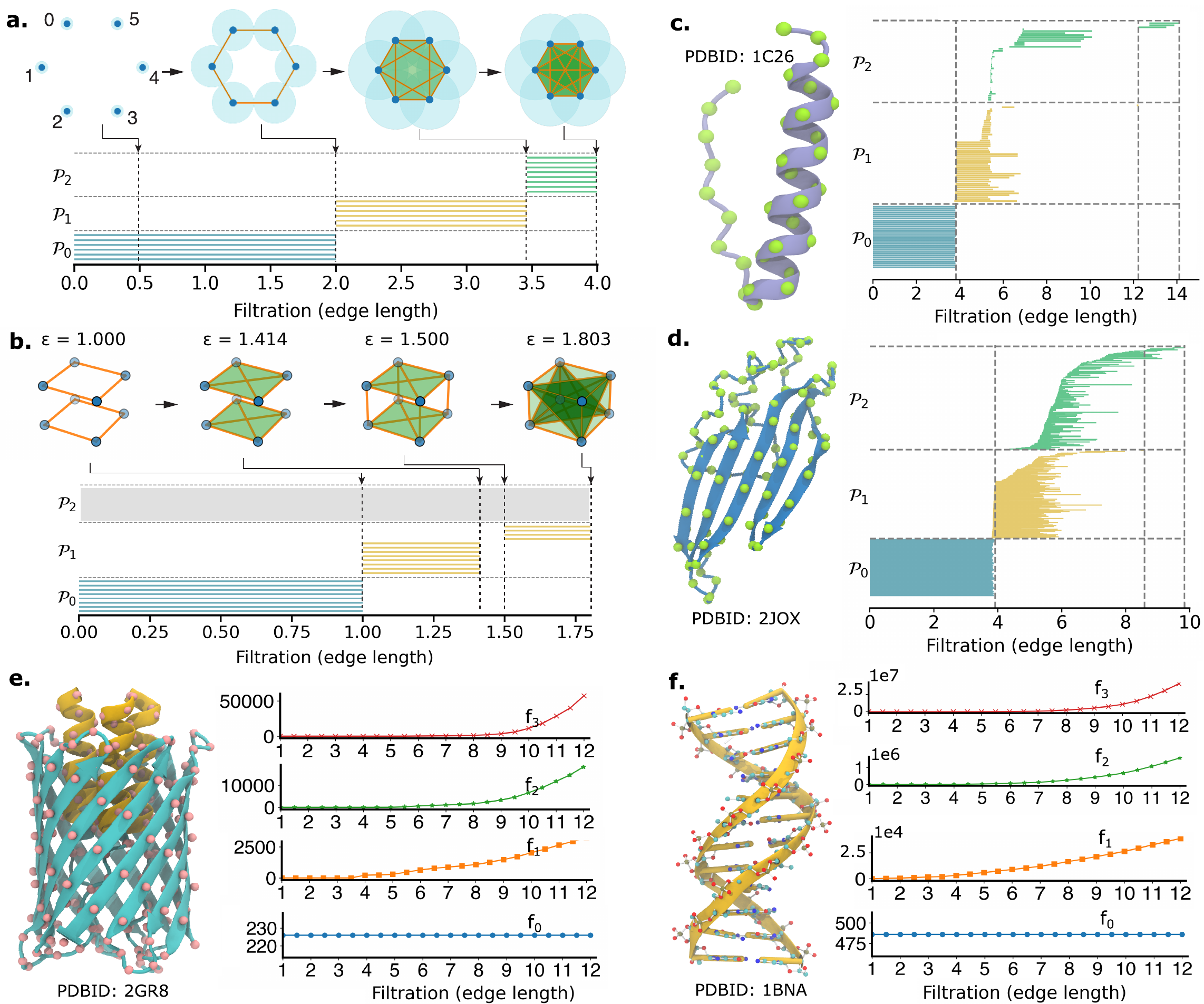}
	\caption{Illustrations of multiscale commutative algebra analysis on point-cloud data using a Rips complex-based filtration process: a. Facet persistence barcode for 6 points. b. Facet persistence barcode for a cuboid with dimensions $1 \times 1 \times 1.5$. c. Facet persistence barcode for the $C_{\alpha}$ atoms of protein 1C26 with alpha-helix structures. d. Facet persistence barcode for the $C_{\alpha}$ atoms of protein 2JOX with beta-sheet structures. e. $f$-vector curves for the $C_{\alpha}$ atoms of protein 2GR8. f. $f$-vector curves for the atoms in DNA structure 1BNA.}
	\label{Fig:PCA-illustrations}
\end{figure}

To illustrate the structural information captured by commutative algebra, we first perform multiscale analysis on point-cloud data using a Rips complex-based filtration process, as demonstrated in \autoref{Fig:PCA-illustrations}. A six-point configuration together with its persistent facet prime ideals in dimensions 0, 1, and 2 is presented in \autoref{Fig:PCA-illustrations}a. In dimension 0, six facet bars persist until the filtration parameter, or edge length, reaches 2, corresponding to the six initially disconnected points. At this value, the points become connected through edges. At a filtration value of 3.5, these edges form triangular boundaries, leading to the disappearance of the corresponding dimension-1 bars. As the filtration parameter increases from 3.5 to 4, higher-dimensional simplices emerge, and the triangular faces become part of higher-order structures. In the barcode visualization, bars with persistence length equal to 0 are excluded.

\autoref{Fig:PCA-illustrations}b shows the filtration process and corresponding facet persistence bars for a cuboid with dimensions $1 \times 1 \times 1.5$. Similar to the six-point example, the dimension-0 facet bars persist until the filtration parameter reaches 1, where the initially independent points become connected by edges of minimal length 1. Two groups of dimension-1 bars are observed. The first group, consisting of 8 bars, represents features that appear in the initial stage of filtration and disappear in the second stage. The second group, consisting of 4 bars, corresponds to vertical edges that emerge in the third stage and vanish in the fourth stage as they merge into triangular faces. No dimension-2 bars are observed because the relevant triangular faces and enclosing tetrahedra form simultaneously.

Next, we illustrate the interpretability of CANet using fundamental protein structural motifs. \autoref{Fig:PCA-illustrations}c displays the facet persistence bars across three dimensions for the $C_{\alpha}$ atoms of an alpha-helix protein structure (PDB ID: 1C26). The dimension-0 bars indicate that these $C_{\alpha}$ atoms become connected within a filtration value of approximately $4\text{\AA}$, consistent with typical distances between adjacent $C_{\alpha}$ atoms. The dimension-1 bars emerge around $4\text{\AA}$, reflecting edge connections between neighboring $C_{\alpha}$ atoms. Because three consecutive $C_{\alpha}$ atoms along an alpha-helix are spatially close, triangular structures form rapidly, resulting in short-lived dimension-1 features. Most of these bars disappear around $5\text{\AA}$, reflecting the compact geometry of the helix, while a few persist until approximately $6.5\text{\AA}$ due to longer connections involving terminal regions. The dimension-2 bars represent triangular faces. In the alpha-helix region, these triangular faces quickly evolve into tetrahedra, producing short persistence intervals, typically between $5\text{\AA}$ and $6\text{\AA}$. The remaining dimension-2 bars, with death values extending up to $14\text{\AA}$, are associated with tetrahedra formed between terminal atoms and atoms in the helix. Thus, the broader range of births and deaths for dimension-2 features reflects geometric variability between the alpha-helix and terminal regions.

\autoref{Fig:PCA-illustrations}d presents the facet persistence analysis for a beta-sheet protein structure (PDB ID: 2JOX) containing 106 $C_{\alpha}$ atoms. As in the alpha-helix case, most dimension-0 bars disappear around $4\text{\AA}$ due to short distances between neighboring $C_{\alpha}$ atoms. The larger number of atoms produces more edges and triangles, resulting in more persistence bars overall. Most dimension-1 bars appear between $4\text{\AA}$ and $6\text{\AA}$. However, the dimension-2 bars tend to persist longer than those in the alpha-helix case because $C_{\alpha}$ atoms in beta-sheets are less likely to form tetrahedra as rapidly as those in helical structures. This comparison shows that persistent facet prime ideals can distinguish local structural organization between alpha-helices and beta-sheets.

For larger biomolecular structures, direct facet barcode analysis becomes computationally expensive due to the rapid growth in the number of edges, triangles, and tetrahedra. The membrane protein shown in \autoref{Fig:PCA-illustrations}e contains 226 $C_{\alpha}$ atoms, making full persistent facet barcode analysis computationally impractical. As a scalable alternative, the $f$-vector records the number of simplices of each dimension across filtration values. The results show that tens of thousands of tetrahedra form as the filtration radius increases, while the dimension-0 $f$-vector curve remains constant because the number of atoms does not change. A similar situation occurs for the DNA structure (PDB ID: 1BNA) in \autoref{Fig:PCA-illustrations}f, which contains 486 atoms. In this case, the number of triangles and tetrahedra can reach magnitudes of $10^6$-$10^7$ at a filtration radius of $12\text{\AA}$, making facet barcode analysis infeasible. Therefore, $f$-vector analysis provides a scalable representation for large protein and nucleic acid structures, although it sacrifices some detailed persistence information.

The importance of developing a unified model capable of predicting disease-associated mutations, protein stability changes, and solubility alterations is clear. Disease-causing mutations can disrupt physical interactions, perturb electrostatic environments, and destabilize local structural regions, ultimately contributing to pathogenic phenotypes. Amino acid substitutions can also alter protein stability and solubility, further modulating protein function. A key barrier to improving CANet's generalizability is the limited availability of well-annotated variants, especially the large number of variants of uncertain significance (VUS), which creates class imbalance and reduces predictive robustness. Addressing this limitation will require integrating larger and more diverse variant resources, such as ClinVar \cite{landrum2018clinvar}, together with high-quality 3D structural predictions from AlphaFold3.

Recent advances such as AlphaMissense, an AlphaFold-based architecture designed to predict the pathogenicity of missense variants \cite{cheng2023accurate}, demonstrate the value of structural context for variant interpretation. Another important direction is to extend CANet to incorporate nonsynonymous single-nucleotide polymorphisms (SNPs) from PolyPhen-2 \cite{adzhubei2010method}. Reliable performance in this broader setting will require high-quality structural priors from AlphaFold3 so that mutation-induced atomic-level perturbations can be characterized more accurately.

Overall, CANet introduces a biologically grounded and algebraically driven XAI framework for interpretable modeling of mutation-induced structural and electrostatic perturbations in proteins. Compared with persistent homology, a well-established method in topological data analysis (TDA) that captures global features such as loops and cavities, commutative algebra provides enhanced sensitivity to local structural changes and electrostatic perturbations. While persistent homology identifies coarse-grained topological invariants, it may miss subtle but biologically critical interactions. In contrast, commutative algebraic descriptors, including facet prime ideals and $f$-vector features, encode multiscale geometric and chemical information. These capabilities enable CANet to detect fine-grained alterations in protein structure and function, highlighting the strength of commutative algebra as a rigorous and interpretable XAI framework for uncovering the molecular basis of genetic diseases.

\section{Software and Resources}
For the M546 dataset, 62 of the 3D protein structures are obtained from the Protein Data Bank (PDB). The remaining structure, PDB ID: 4ZWJ \cite{kang2015crystal}, is predicted by AlphaFold 3 because the experimental PDB structure contains missing residues. The 3D protein structures for the S2648 and S350 datasets are also obtained from the PDB, while the structures for the PON-Sol2 dataset are obtained from those generated in \cite{wee2024integration}.

Mutant proteins are generated using Jackal \cite{xiang2002jackal}. PQR files containing partial charge information are generated using PDB2PQR \cite{dolinsky2004pdb2pqr} with the AMBER force field for wild-type and mutant proteins. Solvation energy and surface area information are computed using the in-house online software packages ESES \cite{liu2017eses} and MIBPB \cite{chen2011mibpb}. p$K_a$ values are computed using PROPKA \cite{li2005very}. Position-specific scoring matrices (PSSMs) are generated using BLAST+ \cite{johnson2008ncbi} with the nr database. Secondary structure and torsion angle sequence-based features are calculated using SPIDER \cite{heffernan2015improving}. Vietoris-Rips (VR) complexes and Alpha complexes are generated using the GUDHI software library \cite{maria2014gudhi}. The commutative algebra embeddings, including facet prime ideals and $f$-vectors, are then calculated following the procedure described in Methods.

\section*{Data Availability}
All datasets analyzed in this study are publicly available from the sources cited in the manuscript. 3D protein structures are available at https://github.com/ExpectozJJ/CANet. 

\section*{Code Availability}
  
The implementation of the proposed CANet and CATree framework is available at 

https://github.com/ExpectozJJ/CANet including the source code for the methods used for comparison in this study.

\section*{Acknowledgments}
This work was supported in part by NIH grants R01AI164266, and R35GM148196, NSF grant  DMS-2052983, MSU Research Foundation, and Bristol-Myers Squibb 65109.
F.S. thanks King Fahd University of Petroleum and Minerals for their support.

\section*{AI Usage Statement}

The authors used an AI-based language model solely to improve the clarity and grammar of the manuscript. The authors take full responsibility for the study design, mathematical formulations, computational analyses, results, interpretations, and final content of the manuscript.

%%\clearpage
% \printindex

%\section*{Supplementary Information}
%Supplementary Information is available and includes additional details about Persistent Stanley-Reisner Theory, multiscale commutative algebra embeddings and datasets used in this work. Additional results, discussion and feature generation details are also provided. 

\newpage

\appendix

\section{Mathematical Background for Multiscale Commutative Algebra Modeling}
\label{app:mathematical-background}
This appendix provides a nontechnical overview of the mathematical ideas underlying the proposed framework. Its purpose is to explain how local protein structures are converted into multiscale algebraic descriptors and, ultimately, into numerical features suitable for machine-learning prediction.

\subsection{From Protein Structures to Simplicial Complexes}

Proteins can be represented geometrically as collections of atoms in
three-dimensional space. Around a mutation site, the spatial arrangement of
these atoms reflects local physical and chemical interactions that may change when one amino acid is replaced by another. Such changes can alter the local packing, hydrogen-bonding network, electrostatic environment, and hydrophobic organization of a protein, thereby influencing its stability, solubility, or association with disease.

In the present work, atomic coordinates are converted into combinatorial
structures known as simplicial complexes. The atoms are represented as
vertices, pairwise atomic interactions are represented as edges, three-atom
interactions are represented as triangles, and higher-order interactions are represented by higher-dimensional simplices. This construction provides a mathematical representation of both the geometry and connectivity of the local molecular environment.

A simplex is included when its vertices satisfy a prescribed geometric
condition. In a Vietoris-Rips complex, a collection of atoms forms a simplex when all pairwise distances between them are below a selected threshold. In an Alpha complex, simplices are determined from the Delaunay triangulation and retain information about the local geometry of the atomic point cloud. These two constructions provide complementary representations of atomic interactions and molecular shape.

\subsection{Filtrations and Multiscale Representation}

A single distance threshold describes the protein structure at only one
spatial scale. However, molecular interactions occur over different length
scales. Short filtration values emphasize closely spaced atoms and strong local contacts, while larger filtration values reveal broader patterns involving more distant atoms. To capture this multiscale organization, the distance threshold is increased gradually.

Let $\Delta^t$ denote the simplicial complex constructed at filtration level $t$. When $s\leq t$, every simplex present in $\Delta^s$ is also present in
$\Delta^t$, giving the nested sequence
\[
\Delta^s\subseteq\Delta^t.
\]
A collection of nested simplicial complexes of this form is called a
filtration. The filtration records when vertices, edges, triangles, and
higher-dimensional structures appear as the spatial scale increases.
Consequently, it describes how the organization of the local protein
environment evolves from short-range to longer-range interactions.

A structure that appears at one filtration level and remains present over
several subsequent levels is considered persistent. Persistent structures are generally less sensitive to small perturbations in atomic coordinates than structures appearing at only one filtration value. The resulting multiscale representation therefore provides a stable description of the mutation environment.

\subsection{Stanley-Reisner Representation}

The central mathematical idea is to translate geometric information about nearby atoms into algebraic information that can be summarized numerically. Readers do not need the technical details of commutative algebra to follow the construction: the polynomial variables represent atoms, while the algebraic relations record which groups of atoms do or do not interact at a given spatial scale.

Persistent Stanley-Reisner theory associates algebraic objects with the
simplicial complexes appearing throughout the filtration. Let
\[
V=\{x_1,\ldots,x_n\}
\]
be the vertex set of a simplicial complex $\Delta$, and let
\[
S=k[x_1,\ldots,x_n]
\]
be the polynomial ring whose variables correspond to the vertices. A
square-free monomial $x_{i_1}\cdots x_{i_r}$ represents a collection of
vertices. When these vertices do not form a simplex in $\Delta$, the
corresponding monomial is included in the Stanley--Reisner ideal
$I(\Delta)$. The associated Stanley--Reisner ring is defined as
\[
k[\Delta]=S/I(\Delta).
\]

In this way, the simplices present in the complex and the interactions missing from the complex are encoded algebraically. As the filtration parameter increases, additional simplices are introduced. Therefore, the simplicial complexes expand while their corresponding Stanley--Reisner ideals decrease:
\[
\Delta^s\subseteq\Delta^t
\qquad\Longleftrightarrow\qquad
I(\Delta^s)\supseteq I(\Delta^t),
\qquad s\leq t.
\]
The evolution of these ideals provides an algebraic description of the
geometric changes occurring throughout the filtration.

A facet is a simplex that is not contained in any larger simplex of the same complex. Facets therefore represent the maximal interacting groups present at a particular filtration level. Each facet $\sigma$ determines a facet prime ideal
\[
P_{\sigma}=(x_i\mid x_i\notin\sigma),
\]
which is generated by the variables corresponding to vertices outside the
facet. The Stanley-Reisner ideal can be expressed as an intersection of these facet prime ideals. Tracking the facet prime ideals across the filtration provides information about the persistence of maximal interaction patterns.

\subsection{Algebraic and Combinatorial Descriptors}

Persistent Stanley-Reisner theory provides several algebraic and
combinatorial descriptors of a filtered simplicial complex. These descriptors summarize different aspects of its structure and evolution.

For filtration levels $t\leq t'$, the facet persistence Betti number
$\beta_i^{t,t'}$ counts the number of $i$-dimensional facet prime ideals that are present at both levels. A large value indicates that many maximal
$i$-dimensional interaction structures remain unchanged over the interval
$[t,t']$. These values provide an algebraic analogue of persistent topological features.

Graded Betti numbers describe relations among the generators of a
Stanley-Reisner ideal or ring. They contain information beyond the number of connected components or cycles by also recording the sizes of the vertex
subsets supporting these structures. Persistent graded Betti numbers extend
this description across filtration levels and connect the algebraic
representation to the persistent homology of induced subcomplexes through
Hochster's formula.

The $f$-vector provides a direct combinatorial summary of a simplicial complex. At filtration level $t$, the quantity $f_i^t$ denotes the number of $i$-dimensional simplices in $\Delta^t$. Thus, $f_0^t$ counts vertices,
$f_1^t$ counts edges, and $f_2^t$ counts triangles. The associated $h$-vector is obtained from the $f$-vector through an invertible linear transformation and provides an alternative representation of the same face-count information.

The facet-based descriptors and $f$-vectors provide complementary
information. The $f$-vector records how many simplices are present, whereas the facet-based descriptors characterize the maximal structures and their persistence. Evaluating these quantities at successive filtration values produces multiscale algebraic and combinatorial curves. In the present study, dimension-0 and dimension-1 facet persistence descriptors and $f$-vectors are used to construct the commutative algebra embeddings.

\subsection{Mutation-Centered Multiscale Vectorization}

The commutative algebra representation is constructed locally around each
mutation site. Let $\mathcal{A}_m$ denote the set of atoms belonging to the
mutated residue, and let $\mathcal{A}_{mn}(r)$ denote the set of neighboring atoms located within a cutoff distance $r$ of the mutation site. In this study, a cutoff distance of $16\,\text{\AA}$ is used to define the mutation neighborhood.

The atoms are further divided into element-specific subsets
$\mathcal{A}_{\varepsilon}$, where
\[
\varepsilon\in\{C,N,O\}.
\]
Combining the three mutation-site element types with the three neighborhood
element types produces nine element-pair representations. These
representations distinguish different chemical environments. For example,
C-C interactions primarily describe hydrophobic contacts, while N-O
interactions reflect hydrophilic contacts that may be associated with hydrogen bonds or salt bridges.

For the Vietoris-Rips construction, a modified Euclidean distance is used to focus specifically on interactions between the mutation site and its
surrounding environment:
\[
{\rm D}_{\mathrm{mod}}(a_i,a_j)=
\begin{cases}
\infty,
& a_i,a_j\in\mathcal{A}_m
\text{ or }
a_i,a_j\in\mathcal{A}_{mn}(r),\\
{\rm DE}(a_i,a_j),
& \text{otherwise},
\end{cases}
\]
where ${\rm DE}(a_i,a_j)$ is the standard Euclidean distance between atoms
$a_i$ and $a_j$. Assigning an infinite distance to pairs belonging entirely to the mutation site or entirely to the neighborhood prevents these pairs from forming edges at finite filtration values. Consequently, the resulting Vietoris-Rips complexes retain only cross interactions between mutation-site atoms and neighborhood atoms. The Alpha complexes are constructed using the standard Euclidean distance and provide complementary geometric information.

The filtration parameter ranges from $1\,\text{\AA}$ to $12\,\text{\AA}$ with an increment of
$0.4\,\text{\AA}$. At each filtration value, dimension-0 and
dimension-1 facet persistence descriptors and $f$-vectors are computed. Their values across all filtration levels form multiscale curves describing the evolution of the local molecular structure.

The complete procedure is applied separately to the wild-type and mutant
protein structures. For each descriptor, the wild-type values, mutant values, and their corresponding differences are included. The difference features directly describe mutation-induced changes in the local atomic organization. The resulting curves are concatenated across filtration levels, dimensions, element-pair combinations, complex types, and structural states to form a fixed-length numerical representation.

The overall construction can be summarized as
\[
\begin{aligned}
\text{protein structure}
&\longrightarrow
\text{mutation-centered atomic point cloud}\\
&\longrightarrow
\text{element-specific filtered simplicial complexes}\\
&\longrightarrow
\text{facet-based and combinatorial descriptors}\\
&\longrightarrow
\text{wild-type, mutant, and difference feature vectors}\\
&\longrightarrow
\text{CANet or CATree prediction}.
\end{aligned}
\]
This representation converts variable-size molecular structures into
fixed-length feature vectors while preserving multiscale information about
local geometry, atomic interactions, and mutation-induced structural changes.
In biological terms, the resulting feature vector summarizes how a mutation changes the organization of nearby atoms across several spatial scales; in computational terms, it provides a consistent numerical input for CANet and CATree.

\subsection{Summary of Mathematical Notation}

Table~\ref{table:appendix-notation} summarizes the principal mathematical and
molecular notation used in the multiscale commutative algebra framework.

\begin{table}[htb!]
    \centering
    \small
    \renewcommand{\arraystretch}{1.15}
    \caption{Principal notation used in the multiscale commutative algebra
    framework.}
    \label{table:appendix-notation}
    \begin{tabular}{p{0.21\textwidth} p{0.69\textwidth}}
        \hline
        \textbf{Symbol} & \textbf{Meaning} \\
        \hline

        $V=\{x_1,\ldots,x_n\}$ &
        Finite vertex set representing the atoms included in a molecular
        structure. \\

        $\Delta$ &
        Simplicial complex constructed from an atomic point cloud. \\

        $\sigma$ &
        A simplex or face of $\Delta$, such as a vertex, edge, or triangle. \\

        $t$ &
        Filtration parameter controlling the spatial scale at which atomic
        interactions are included. \\

        $\Delta^t$ &
        Simplicial complex constructed at filtration level $t$. \\

        $S=k[x_1,\ldots,x_n]$ &
        Polynomial ring whose variables correspond to the vertices of the
        simplicial complex. \\

        $I^t=I(\Delta^t)$ &
        Stanley-Reisner ideal encoding vertex subsets that do not form
        simplices in $\Delta^t$. \\

        $k[\Delta^t]=S/I^t$ &
        Stanley-Reisner ring associated with the filtered simplicial complex. \\

        $\mathcal{F}(\Delta^t)$ &
        Set of facets, or maximal simplices, of $\Delta^t$. \\

        $P_{\sigma}$ &
        Facet prime ideal associated with the facet $\sigma$. \\

        $\mathcal{P}_i^t$ &
        Collection of facet prime ideals associated with $i$-dimensional
        facets at filtration level $t$. \\

        $\beta_i^{t,t'}$ &
        Number of $i$-dimensional facet prime ideals persisting from level
        $t$ to level $t'$. \\

        $\beta_{i,j}^{t,t'}$ &
        Persistent graded Betti number in homological degree $i$ and internal
        degree $j$. \\

        $f_i^t$ &
        Number of $i$-dimensional simplices at filtration level $t$. \\

        $h_i^t$ &
        Combinatorial quantity obtained from the $f$-vector through an
        invertible linear transformation. \\

        $\mathcal{A}_m$ &
        Set of atoms belonging to the mutation site. \\

        $\mathcal{A}_{mn}(r)$ &
        Set of neighboring atoms located within distance $r$ of the mutation
        site. \\

        $\mathcal{A}_{\varepsilon}$ &
        Element-specific atom set, where
        $\varepsilon\in\{C,N,O\}$. \\

        ${\rm DE}$ &
        Standard Euclidean distance between two atoms. \\

        ${\rm D}_{\mathrm{mod}}$ &
        Modified Euclidean distance retaining only interactions between
        mutation-site atoms and neighborhood atoms. \\

        $r$ &
        Cutoff radius used to define the mutation neighborhood. \\

        \hline
    \end{tabular}
\end{table}

\section{Further Reading}

The following references provide background on the mathematical,
computational, and biological ideas underlying the present work.

\subsection*{Mathematical and Computational Background}
\begin{itemize}
    
       \item 
F. Suwayyid  and G.-W. Wei, Persistent Stanley-Reisner theory,  Foundations of Data Science (Springfield, Mo.), 8 (2026), pp. 287-312.

This article introduces the persistent Stanley-Reisner framework used in the
present work and explains how commutative-algebraic invariants evolve across a
filtration.

  \item 
F. Suwayyid,  Y. Hozumi,  M. Zia, H. Feng, J. Wee, and G.-W. Wei,  CAKR: Commutative algebra $k$-mer representations for genomics,
Nature Communications, in press (2026).  

This work illustrates another application of commutative algebra to biological
data and shows how algebraic representations can be integrated with machine
learning.
\end{itemize}

\subsection*{Biological Background}
 
 \begin{itemize}
        \item 
Z.E. Sauna and C. Kimchi-Sarfaty, Understanding the contribution of synonymous mutations to human disease, \emph{Nature Reviews Genetics}, 12 (2011), pp.~683--691.

This review describes molecular mechanisms through which genetic variants can
influence gene expression, protein folding, and human disease.

\end{itemize}

%\clearpage

\bibliographystyle{unsrt}
\bibliography{refs}

@article{xiang2002jackal,
  title={Jackal: {A} protein structure modeling package},
  author={Xiang, Jason Z and Honig, B},
  journal={Columbia University and Howard Hughes Medical Institute, New York},
  year={2002}
}

@article{adzhubei2010method,
  title={A method and server for predicting damaging missense mutations},
  author={Adzhubei, Ivan A and Schmidt, Steffen and Peshkin, Leonid and Ramensky, Vasily E and Gerasimova, Anna and Bork, Peer and Kondrashov, Alexey S and Sunyaev, Shamil R},
  journal={Nature methods},
  volume={7},
  number={4},
  pages={248--249},
  year={2010},
  publisher={Nature Publishing Group US New York}
}

@article{landrum2018clinvar,
  title={Clin{V}ar: improving access to variant interpretations and supporting evidence},
  author={Landrum, Melissa J and Lee, Jennifer M and Benson, Mark and Brown, Garth R and Chao, Chen and Chitipiralla, Shanmuga and Gu, Baoshan and Hart, Jennifer and Hoffman, Douglas and Jang, Wonhee and others},
  journal={Nucleic acids research},
  volume={46},
  number={D1},
  pages={D1062--D1067},
  year={2018},
  publisher={Oxford University Press}
}

@article{cheng2023accurate,
  title={Accurate proteome-wide missense variant effect prediction with {AlphaM}issense},
  author={Cheng, Jun and Novati, Guido and Pan, Joshua and Bycroft, Clare and {\v{Z}}emgulyt{\.e}, Akvil{\.e} and Applebaum, Taylor and Pritzel, Alexander and Wong, Lai Hong and Zielinski, Michal and Sargeant, Tobias and others},
  journal={Science},
  volume={381},
  number={6664},
  pages={eadg7492},
  year={2023},
  publisher={American Association for the Advancement of Science}
}

@article{haass2007soluble,
	title={Soluble protein oligomers in neurodegeneration: lessons from the Alzheimer's amyloid $\beta$-peptide},
	author={Haass, Christian and Selkoe, Dennis J},
	journal={Nature reviews Molecular cell biology},
	volume={8},
	number={2},
	pages={101--112},
	year={2007},
	publisher={Nature Publishing Group UK London}
}

@article{suwayyid2025persistent,
title = {Persistent {Stanley–Reisner} theory},
author = {Faisal Suwayyid and Guo-Wei Wei},
journal = {Foundations of Data Science},
pages = {},
year = {2025},
issn = {},
doi = {10.3934/fods.2025009},
url = {https://www.aimsciences.org/article/id/68480494b924fa00d9d4efaa}
}

@article{lin2023evolutionary,
  title={Evolutionary-scale prediction of atomic-level protein structure with a language model},
  author={Lin, Zeming and Akin, Halil and Rao, Roshan and Hie, Brian and Zhu, Zhongkai and Lu, Wenting and Smetanin, Nikita and Verkuil, Robert and Kabeli, Ori and Shmueli, Yaniv and others},
  journal={Science},
  volume={379},
  number={6637},
  pages={1123--1130},
  year={2023},
  publisher={American Association for the Advancement of Science}
}

@article{ding2022co,
  title={Co-evolution of interacting proteins through non-contacting and non-specific mutations},
  author={Ding, David and Green, Anna G and Wang, Boyuan and Lite, Thuy-Lan Vo and Weinstein, Eli N and Marks, Debora S and Laub, Michael T},
  journal={Nature ecology \& evolution},
  volume={6},
  number={5},
  pages={590--603},
  year={2022},
  publisher={Nature Publishing Group UK London}
}

@article{worth2011sdm,
  title={{SDM}—a server for predicting effects of mutations on protein stability and malfunction},
  author={Worth, Catherine L and Preissner, Robert and Blundell, Tom L},
  journal={Nucleic acids research},
  volume={39},
  number={suppl\_2},
  pages={W215--W222},
  year={2011},
  publisher={Oxford University Press}
}

@article{pires2020mcsm,
  title={m{CSM}-membrane: predicting the effects of mutations on transmembrane proteins},
  author={Pires, Douglas EV and Rodrigues, Carlos HM and Ascher, David B},
  journal={Nucleic Acids Research},
  volume={48},
  number={W1},
  pages={W147--W153},
  year={2020},
  publisher={Oxford University Press}
}

@article{karczewski2020mutational,
  title={The mutational constraint spectrum quantified from variation in 141,456 humans},
  author={Karczewski, Konrad J and Francioli, Laurent C and Tiao, Grace and Cummings, Beryl B and Alf{\"o}ldi, Jessica and Wang, Qingbo and Collins, Ryan L and Laricchia, Kristen M and Ganna, Andrea and Birnbaum, Daniel P and others},
  journal={Nature},
  volume={581},
  number={7809},
  pages={434--443},
  year={2020},
  publisher={Nature Publishing Group UK London}
}

@article{wee2024integration,
  title={Integration of persistent {L}aplacian and pre-trained transformer for protein solubility changes upon mutation},
  author={Wee, JunJie and Chen, Jiahui and Xia, Kelin and Wei, Guo-Wei},
  journal={Computers in biology and medicine},
  volume={169},
  pages={107918},
  year={2024},
  publisher={Elsevier}
}

@article{cang2017topologynet,
  title={Topology{Net: T}opology based deep convolutional and multi-task neural networks for biomolecular property predictions},
  author={Cang, Zixuan and Wei, Guo-Wei},
  journal={PLoS computational biology},
  volume={13},
  number={7},
  pages={e1005690},
  year={2017},
  publisher={Public Library of Science San Francisco, CA USA}
}

@article{yang2021pon,
  title={{PON-S}ol2: prediction of effects of variants on protein solubility},
  author={Yang, Yang and Zeng, Lianjie and Vihinen, Mauno},
  journal={International Journal of Molecular Sciences},
  volume={22},
  number={15},
  pages={8027},
  year={2021},
  publisher={MDPI}
}

@article{popov2019prediction,
  title={Prediction of disease-associated mutations in the transmembrane regions of proteins with known 3{D} structure},
  author={Popov, Petr and Bizin, Ilya and Gromiha, Michael and A, Kulandaisamy and Frishman, Dmitrij},
  journal={PloS one},
  volume={14},
  number={7},
  pages={e0219452},
  year={2019},
  publisher={Public Library of Science San Francisco, CA USA}
}

@article{levy2010simple,
  title={A simple definition of structural regions in proteins and its use in analyzing interface evolution},
  author={Levy, Emmanuel D},
  journal={Journal of molecular biology},
  volume={403},
  number={4},
  pages={660--670},
  year={2010},
  publisher={Elsevier}
}

@article{bogan1998anatomy,
  title={Anatomy of hot spots in protein interfaces},
  author={Bogan, Andrew A and Thorn, Kurt S},
  journal={Journal of molecular biology},
  volume={280},
  number={1},
  pages={1--9},
  year={1998},
  publisher={Elsevier}
}

@article{quan2016strum,
  title={{STRUM}: structure-based prediction of protein stability changes upon single-point mutation},
  author={Quan, Lijun and Lv, Qiang and Zhang, Yang},
  journal={Bioinformatics},
  volume={32},
  number={19},
  pages={2936--2946},
  year={2016},
  publisher={Oxford University Press}
}

@article{dehouck2009fast,
  title={Fast and accurate predictions of protein stability changes upon mutations using statistical potentials and neural networks: {PoPMuSiC}-2.0},
  author={Dehouck, Yves and Grosfils, Aline and Folch, Benjamin and Gilis, Dimitri and Bogaerts, Philippe and Rooman, Marianne},
  journal={Bioinformatics},
  volume={25},
  number={19},
  pages={2537--2543},
  year={2009},
  publisher={Oxford University Press}
}

@article{sun2022electrostatics,
  title={Electrostatics in computational biophysics and its implications for disease effects},
  author={Sun, Shengjie and Poudel, Pitambar and Alexov, Emil and Li, Lin},
  journal={International Journal of Molecular Sciences},
  volume={23},
  number={18},
  pages={10347},
  year={2022},
  publisher={MDPI}
}

@article{wang2021genetic,
  title={Genetic prion disease--related mutation {E196K} displays a novel amyloid fibril structure revealed by cryo-{EM}},
  author={Wang, Li-Qiang and Zhao, Kun and Yuan, Han-Ye and Li, Xiang-Ning and Dang, Hai-Bin and Ma, Yeyang and Wang, Qiang and Wang, Chen and Sun, Yunpeng and Chen, Jie and others},
  journal={Science advances},
  volume={7},
  number={37},
  pages={eabg9676},
  year={2021},
  publisher={American Association for the Advancement of Science}
}

@article{wrapp2020cryo,
  title={{Cryo-EM} structure of the 2019-{nCoV} spike in the prefusion conformation},
  author={Wrapp, Daniel and Wang, Nianshuang and Corbett, Kizzmekia S and Goldsmith, Jory A and Hsieh, Ching-Lin and Abiona, Olubukola and Graham, Barney S and McLellan, Jason S},
  journal={Science},
  volume={367},
  number={6483},
  pages={1260--1263},
  year={2020},
  publisher={American Association for the Advancement of Science}
}

@article{scikit-learn,
 title={Scikit-learn: {Machine Learning} in {P}ython},
 author={Pedregosa, F. and Varoquaux, G. and Gramfort, A. and Michel, V.
         and Thirion, B. and Grisel, O. and Blondel, M. and Prettenhofer, P.
         and Weiss, R. and Dubourg, V. and Vanderplas, J. and Passos, A. and
         Cournapeau, D. and Brucher, M. and Perrot, M. and Duchesnay, E.},
 journal={Journal of Machine Learning Research},
 volume={12},
 pages={2825--2830},
 year={2011}
}

@article{bas2008very,
  title={Very fast prediction and rationalization of p{K}a values for protein--ligand complexes},
  author={Bas, Delphine C and Rogers, David M and Jensen, Jan H},
  journal={Proteins: Structure, Function, and Bioinformatics},
  volume={73},
  number={3},
  pages={765--783},
  year={2008},
  publisher={Wiley Online Library}
}

@inproceedings{maria2014gudhi,
  title={The {GUDHI} library: {S}implicial complexes and persistent homology},
  author={Maria, Cl{\'e}ment and Boissonnat, Jean-Daniel and Glisse, Marc and Yvinec, Mariette},
  booktitle={Mathematical Software--ICMS 2014: 4th International Congress, Seoul, South Korea, August 5-9, 2014. Proceedings 4},
  pages={167--174},
  year={2014},
  organization={Springer}
}

@article{heffernan2015improving,
  title={Improving prediction of secondary structure, local backbone angles and solvent accessible surface area of proteins by iterative deep learning},
  author={Heffernan, Rhys and Paliwal, Kuldip and Lyons, James and Dehzangi, Abdollah and Sharma, Alok and Wang, Jihua and Sattar, Abdul and Yang, Yuedong and Zhou, Yaoqi},
  journal={Scientific reports},
  volume={5},
  number={1},
  pages={11476},
  year={2015},
  publisher={Nature Publishing Group UK London}
}

@article{johnson2008ncbi,
  title={{NCBI BLAST}: {A} better web interface},
  author={Johnson, Mark and Zaretskaya, Irena and Raytselis, Yan and Merezhuk, Yuri and McGinnis, Scott and Madden, Thomas L},
  journal={Nucleic acids research},
  volume={36},
  number={suppl\_2},
  pages={W5--W9},
  year={2008},
  publisher={Oxford University Press}
}

@article{li2005very,
  title={Very fast empirical prediction and rationalization of protein {pKa} values},
  author={Li, Hui and Robertson, Andrew D and Jensen, Jan H},
  journal={Proteins: Structure, Function, and Bioinformatics},
  volume={61},
  number={4},
  pages={704--721},
  year={2005},
  publisher={Wiley Online Library}
}

@article{chen2011mibpb,
  title={M{IBPB}: {A} software package for electrostatic analysis},
  author={Chen, Duan and Chen, Zhan and Chen, Changjun and Geng, Weihua and Wei, Guo-Wei},
  journal={Journal of computational chemistry},
  volume={32},
  number={4},
  pages={756--770},
  year={2011},
  publisher={Wiley Online Library}
}

@article{adzhubei2010polyphen,
  title={A method and server for predicting damaging missense mutations},
  author={Adzhubei, Ivan A and Schmidt, Steffen and Peshkin, Leonid and Ramensky, Vasily E and Gerasimova, Anna and Bork, Peer and Kondrashov, Alexey S and Sunyaev, Shamil R},
  journal={Nature Methods},
  volume={7},
  number={4},
  pages={248--249},
  year={2010},
  publisher={Nature Publishing Group},
  doi={10.1038/nmeth0410-248}
}

@article{kang2015crystal,
  title={Crystal structure of rhodopsin bound to arrestin by femtosecond {X}-ray laser},
  author={Kang, Yanyong and Zhou, X Edward and Gao, Xiang and He, Yuanzheng and Liu, Wei and Ishchenko, Andrii and Barty, Anton and White, Thomas A and Yefanov, Oleksandr and Han, Gye Won and others},
  journal={Nature},
  volume={523},
  number={7562},
  pages={561--567},
  year={2015},
  publisher={Nature Publishing Group UK London}
}

@article{ng2003sift,
  title={S{IFT}: {P}redicting amino acid changes that affect protein function},
  author={Ng, Pauline C and Henikoff, Steven},
  journal={Nucleic Acids Research},
  volume={31},
  number={13},
  pages={3812--3814},
  year={2003},
  doi={10.1093/nar/gkg509}
}

@article{kircher2014cadd,
  title={A general framework for estimating the relative pathogenicity of human genetic variants},
  author={Kircher, Martin and Witten, Daniela M and Jain, Preti and O'Roak, Brian J and Cooper, Gregory M and Shendure, Jay},
  journal={Nature Genetics},
  volume={46},
  number={3},
  pages={310--315},
  year={2014},
  doi={10.1038/ng.2892}
}

@inproceedings{Devlin2019BERTPO,
  title={B{ERT}: {Pre-training of Deep Bidirectional Transformers for Language Understanding}},
  author={Jacob Devlin and Ming-Wei Chang and Kenton Lee and Kristina Toutanova},
  booktitle={North American Chapter of the Association for Computational Linguistics},
  year={2019},
  url={https://api.semanticscholar.org/CorpusID:52967399}
}

@article{calvanese2018structural,
  title={Structural basis for mutations of human aquaporins associated to genetic diseases},
  author={Calvanese, Luisa and D’Auria, Gabriella and Vangone, Anna and Falcigno, Lucia and Oliva, Romina},
  journal={International Journal of Molecular Sciences},
  volume={19},
  number={6},
  pages={1577},
  year={2018},
  publisher={MDPI}
}

@article{molnar2016characterization,
  title={Characterization of disease-associated mutations in human transmembrane proteins},
  author={Moln{\'a}r, J{\'a}nos and Szak{\'a}cs, Gergely and Tusn{\'a}dy, G{\'a}bor E},
  journal={PloS one},
  volume={11},
  number={3},
  pages={e0151760},
  year={2016},
  publisher={Public Library of Science San Francisco, CA USA}
}

@article{vihinen2020solubility,
  title={Solubility of proteins},
  author={Vihinen, Mauno},
  journal={ADMET and DMPK},
  volume={8},
  number={4},
  pages={391--399},
  year={2020},
  publisher={Me{\dj}unarodna udruga fizikalnih kemi{\v{c}}ara}
}

@article{van2016solubis,
  title={Solubis: a webserver to reduce protein aggregation through mutation},
  author={Van Durme, Joost and De Baets, Greet and Van Der Kant, Rob and Ramakers, Meine and Ganesan, Ashok and Wilkinson, Hannah and Gallardo, Rodrigo and Rousseau, Frederic and Schymkowitz, Joost},
  journal={Protein Engineering, Design and Selection},
  volume={29},
  number={8},
  pages={285--289},
  year={2016},
  publisher={Oxford University Press}
}

@article{nakaya2018amyotrophic,
  title={Amyotrophic {L}ateral {S}clerosis associated {FUS} mutation shortens mitochondria and induces neurotoxicity},
  author={Nakaya, Tadashi and Maragkakis, Manolis},
  journal={Scientific reports},
  volume={8},
  number={1},
  pages={15575},
  year={2018},
  publisher={Nature Publishing Group UK London}
}

@article{funayama2023molecular,
  title={Molecular genetics of {P}arkinson’s disease: {C}ontributions and global trends},
  author={Funayama, Manabu and Nishioka, Kenya and Li, Yuanzhe and Hattori, Nobutaka},
  journal={Journal of Human Genetics},
  volume={68},
  number={3},
  pages={125--130},
  year={2023},
  publisher={Springer Nature Singapore Singapore}
}

@article{hurley2023familial,
  title={Familial {A}lzheimer’s disease-associated {PSEN}1 mutations affect neurodevelopment through increased {N}otch signaling},
  author={Hurley, Erin M and Mozolewski, Pawel and Dobrowolski, Radek and Hsieh, Jenny},
  journal={Stem cell reports},
  volume={18},
  number={7},
  pages={1516--1533},
  year={2023},
  publisher={Elsevier}
}

@article{chen2022mutant,
  title={Mutant p53 in cancer: from molecular mechanism to therapeutic modulation},
  author={Chen, Xiaohua and Zhang, Taotao and Su, Wei and Dou, Zhihui and Zhao, Dapeng and Jin, Xiaodong and Lei, Huiwen and Wang, Jing and Xie, Xiaodong and Cheng, Bo and others},
  journal={Cell death \& disease},
  volume={13},
  number={11},
  pages={974},
  year={2022},
  publisher={Nature Publishing Group UK London}
}

@article{li2024prediction,
  title={Prediction of mutation-induced protein stability changes based on the geometric representations learned by a self-supervised method},
  author={Li, Shan Shan and Liu, Zhao Ming and Li, Jiao and Ma, Yi Bo and Dong, Ze Yuan and Hou, Jun Wei and Shen, Fu Jie and Wang, Wei Bu and Li, Qi Ming and Su, Ji Guo},
  journal={BMC bioinformatics},
  volume={25},
  number={1},
  pages={282},
  year={2024},
  publisher={Springer}
}

@article{stefl2013molecular,
  title={Molecular mechanisms of disease-causing missense mutations},
  author={Stefl, Shannon and Nishi, Hafumi and Petukh, Marharyta and Panchenko, Anna R and Alexov, Emil},
  journal={Journal of molecular biology},
  volume={425},
  number={21},
  pages={3919--3936},
  year={2013},
  publisher={Elsevier}
}

@article{zhang2013y328c,
  title={A {Y328C} missense mutation in spermine synthase causes a mild form of {Snyder--R}obinson syndrome},
  author={Zhang, Zhe and Norris, Joy and Kalscheuer, Vera and Wood, Tim and Wang, Lin and Schwartz, Charles and Alexov, Emil and Van Esch, Hilde},
  journal={Human molecular genetics},
  volume={22},
  number={18},
  pages={3789--3797},
  year={2013},
  publisher={Oxford University Press}
}

@article{zhang2011silico,
  title={In silico and in vitro investigations of the mutability of disease-causing missense mutation sites in spermine synthase},
  author={Zhang, Zhe and Norris, Joy and Schwartz, Charles and Alexov, Emil},
  journal={PloS one},
  volume={6},
  number={5},
  pages={e20373},
  year={2011},
  publisher={Public Library of Science San Francisco, USA}
}

@article{zhang2013rational,
  title={A rational free energy-based approach to understanding and targeting disease-causing missense mutations},
  author={Zhang, Zhe and Witham, Shawn and Petukh, Marharita and Moroy, Gautier and Miteva, Maria and Ikeguchi, Yoshihiko and Alexov, Emil},
  journal={Journal of the American Medical Informatics Association},
  volume={20},
  number={4},
  pages={643--651},
  year={2013},
  publisher={BMJ Publishing Group BMA House, Tavistock Square, London, WC1H 9JR}
}

@article{yang2016pon,
  title={P{ON-S}ol: {P}rediction of effects of amino acid substitutions on protein solubility},
  author={Yang, Yang and Niroula, Abhishek and Shen, Bairong and Vihinen, Mauno},
  journal={Bioinformatics},
  volume={32},
  number={13},
  pages={2032--2034},
  year={2016},
  publisher={Oxford University Press}
}

@article{paladin2017soda,
  title={{SODA}: {P}rediction of protein solubility from disorder and aggregation propensity},
  author={Paladin, Lisanna and Piovesan, Damiano and Tosatto, Silvio CE},
  journal={Nucleic acids research},
  volume={45},
  number={W1},
  pages={W236--W240},
  year={2017},
  publisher={Oxford University Press}
}

@article{sormanni2015camsol,
  title={The {CamSol} method of rational design of protein mutants with enhanced solubility},
  author={Sormanni, Pietro and Aprile, Francesco A and Vendruscolo, Michele},
  journal={Journal of molecular biology},
  volume={427},
  number={2},
  pages={478--490},
  year={2015},
  publisher={Elsevier}
}

@article{vaswani2017attention,
  title={Attention is all you need},
  author={Vaswani, Ashish and Shazeer, Noam and Parmar, Niki and Uszkoreit, Jakob and Jones, Llion and Gomez, Aidan N and Kaiser, {\L}ukasz and Polosukhin, Illia},
  journal={Advances in {N}eural {I}nformation {P}rocessing {S}ystems},
  volume={30},
  year={2017}
}

@article{molotkov2024making,
  title={Making sense of missense: challenges and opportunities in variant pathogenicity prediction},
  author={Molotkov, Ivan and Mardis, Elaine R and Artomov, Mykyta},
  journal={Disease Models \& Mechanisms},
  volume={17},
  number={12},
  pages={dmm052218},
  year={2024},
  publisher={The Company of Biologists Ltd}
}

@article{marian2020clinical,
  title={Clinical interpretation and management of genetic variants},
  author={Marian, Ali J},
  journal={Basic to Translational Science},
  volume={5},
  number={10},
  pages={1029--1042},
  year={2020},
  publisher={American College of Cardiology Foundation Washington DC}
}

@article{niroula2015pon,
  title={{PON-P2}: prediction method for fast and reliable identification of harmful variants},
  author={Niroula, Abhishek and Urolagin, Siddhaling and Vihinen, Mauno},
  journal={PloS one},
  volume={10},
  number={2},
  pages={e0117380},
  year={2015},
  publisher={Public Library of Science San Francisco, CA USA}
}

@misc{liu2017eses,
  title={{ESES}: {S}oftware for eulerian solvent excluded surface},
  author={Liu, Beibei and Wang, Bao and Zhao, Rundong and Tong, Yiying and Wei, Guo-Wei},
  year={2017},
  publisher={Wiley Online Library}
}

@article{dolinsky2004pdb2pqr,
  title={{PDB2PQR}: {A}n automated pipeline for the setup of {Poisson--B}oltzmann electrostatics calculations},
  author={Dolinsky, Todd J and Nielsen, Jens E and McCammon, J Andrew and Baker, Nathan A},
  journal={Nucleic acids research},
  volume={32},
  number={suppl\_2},
  pages={W665--W667},
  year={2004},
  publisher={Oxford University Press}
}

@article{plante2021spike,
  title={Spike mutation {D614G alters SARS-CoV-2} fitness},
  author={Plante, Jessica A and Liu, Yang and Liu, Jianying and Xia, Hongjie and Johnson, Bryan A and Lokugamage, Kumari G and Zhang, Xianwen and Muruato, Antonio E and Zou, Jing and Fontes-Garfias, Camila R and others},
  journal={Nature},
  volume={592},
  number={7852},
  pages={116--121},
  year={2021},
  publisher={Nature Publishing Group UK London}
}

@article{wang2020topology,
  title={A topology-based network tree for the prediction of protein--protein binding affinity changes following mutation},
  author={Wang, Menglun and Cang, Zixuan and Wei, Guo-Wei},
  journal={Nature Machine Intelligence},
  volume={2},
  number={2},
  pages={116--123},
  year={2020},
  publisher={Nature Publishing Group UK London}
}

@article{chen2023topological,
  title={Topological deep learning based deep mutational scanning},
  author={Chen, Jiahui and Woldring, Daniel R and Huang, Faqing and Huang, Xuefei and Wei, Guo-Wei},
  journal={Computers in biology and medicine},
  volume={164},
  pages={107258},
  year={2023},
  publisher={Elsevier}
}

@book{eisenbud2013commutative,
  title={Commutative algebra: with a view toward algebraic geometry},
  author={Eisenbud, David},
  volume={150},
  year={2013},
  publisher={Springer Science \& Business Media}
}

@book{miller2005combinatorial,
	title     = {Combinatorial Commutative Algebra},
	author    = {Miller, Ezra and Sturmfels, Bernd},
	publisher = {Springer},
	year      = {2005},
	series    = {Graduate Texts in Mathematics},
	volume    = {227},
	isbn      = {978-0-387-22356-8},
	doi       = {10.1007/b138602},
	url       = {https://link.springer.com/book/10.1007/b138602}
}

@article{rives2021biological,
	title={Biological structure and function emerge from scaling unsupervised learning to 250 million protein sequences},
	author={Rives, Alexander and Meier, Joshua and Sercu, Tom and Goyal, Siddharth and Lin, Zeming and Liu, Jason and Guo, Demi and Ott, Myle and Zitnick, C Lawrence and Ma, Jerry and others},
	journal={Proceedings of the National Academy of Sciences},
	volume={118},
	number={15},
	pages={e2016239118},
	year={2021},
	publisher={National Acad Sciences}
}
\end{document}